\documentclass[longauth]{aa}
\usepackage{txfonts}
\usepackage{graphicx}
\usepackage{amsmath}
\usepackage{mathrsfs}
\usepackage[breaklinks, colorlinks, citecolor=blue, urlcolor = blue]{hyperref}
\usepackage{pdflscape}
\usepackage{float}
\usepackage{caption}
\usepackage{here}
\usepackage{xcolor}
\usepackage[export]{adjustbox}
\usepackage{array}
\usepackage{natbib}
\usepackage{multirow}
\usepackage{arydshln}
\usepackage{placeins}
\bibpunct{(}{)}{;}{a}{}{,}
\newcommand{\xmm}{\textit{XMM-Newton}}
\newcommand{\mseed}{$M_{\rm s, Edd}$}

\makeatletter
\newsavebox\sidefigbox
\newlength\sidecapht
\long\def\aa@makesidecaption#1#2{%
  \parbox[b][\sidecapht][t]{\aa@sidecaptionwidth}{%
    \aa@captionfont
    \aa@@makecaption{#1}
    \aa@capstrut
    #2%
  }%
}
\makeatother
\defcitealias{Zappacosta2023}{Z23}
\defcitealias{Tortosa2024}{T24}

\begin{document} 
   \title{Widespread steep X-ray spectra of luminous $z\simeq5.85-7.6$ quasars and their implications for future X-ray surveys.}
   \titlerunning{Steep X-ray spectra of EoR QSOs}
   \authorrunning{Tortosa et al., 2026}

   \author{A. Tortosa\thanks{\email{alessia.tortosa@inaf.it}}\inst{1,2}
          \and L. Zappacosta \inst{2}
          \and E. Piconcelli \inst{2}
          \and I. Saccheo \inst{3}
          \and G. Lanzuisi \inst{1}
          \and E. Bertola \inst{4}
          \and F. Vito \inst{1}
          \and M. Bischetti \inst{5,6,7}
          \and M. Brazzini \inst{8,6,7}
          \and M. Castellano \inst{2}
          \and S. Carniani \inst{9}
          \and A. Comastri \inst{1}
          \and S. Gallerani \inst{9}
          \and R. Gilli \inst{1}
          \and F. La Franca \inst{10,2}
          \and P. Madau \inst{11}
          \and R. Maiolino \inst{12,13,14}
          \and G. Miniutti \inst{15}
          \and B. Musiimenta \inst{2}
          \and C. Piscitelli \inst{8,6}
          \and F. Salvestrini \inst{6,7}
          \and F. Tombesi \inst{16,17,2}
          \and R. Tripodi \inst{2}
          \and C. Vignali \inst{1}
          \and M. Volonteri \inst{18}
          \and A. Bongiorno \inst{2}
          \and L. Borrelli \inst{19,1}
          \and M. Brusa \inst{19,1}
          \and I. V. Chilingarian \inst{20,21} 
          \and F. Civano \inst{22}
          \and S. Cristiani \inst{6,7,23}
          \and V. D'Odorico \inst{6,7}
          \and D. De Cicco \inst{24,25}
          \and C. Done \inst{26}
          \and M. Elvis \inst{27}
          \and X. Fan \inst{28}
          \and C. Feruglio \inst{6,7}
          \and F. Fiore \inst{6,7}
          \and A. Grazian \inst{29}
          \and M. Guainazzi \inst{30}
          \and F. Haardt \inst{31,32,33}
          \and A. Luminari \inst{2}
          \and N. Menci \inst{2}
          \and R. Middei \inst{2}
          \and F. Nicastro \inst{2}
          \and M. Paolillo \inst{23,24}
          \and M. Perna \inst{15}
          \and S. Puccetti \inst{34}
          \and R. Schneider \inst{2,35,36,37}
          \and V. Testa \inst{2}
          \and R. Valiante \inst{2,36}
           \and L. Vallini \inst{1}
          \and E. Vanzella \inst{1}
          \and G. Vietri \inst{38}
          }
   \institute{INAF - Osservatorio di Astrofisica e Scienza dello Spazio di Bologna, via Piero Gobetti, 93/3, I-40129 Bologna, Italy
   \and INAF - Osservatorio Astronomico di Roma, via Frascati 33, I-00078 Monte Porzio Catone, Italy.
   \and School of Physics, HH Wills Physics Laboratory, University of Bristol, Tyndall Avenue, Bristol, BS8 1TL, UK
   \and INAF — Osservatorio Astrofisico di Arcetri, Largo E. Fermi 5, I-50125, Florence, Italy.
   \and Dipartimento di Fisica ”Enrico Fermi”, Università di Pisa, Largo Bruno Pontecorvo 3, Pisa, I-56127, Italy
   \and INAF – Osservatorio Astronomico di Trieste, Via G. B. Tiepolo 11, I-34143 Trieste, Italy.%
   \and IFPU - Institute for Fundamental Physics of the Universe, via Beirut 2, I-34151 Trieste, Italy.
   \and Department of Physics, Astronomy Section, University of Trieste, Via G.B. Tiepolo, 11, I-34143 Trieste, Italy.%
   \and Scuola Normale Superiore, Piazza dei Cavalieri 7, I-56126 Pisa, Italy.%
   \and Dipartimento di Matematica e Fisica, Università Roma Tre, Via della Vasca Navale 84, 00146 Roma, Italy.
   \and Department of Astronomy and Astrophysics, University of California, Santa Cruz,California 95064.
   \and Cavendish Laboratory, University of Cambridge, 19 J. J. Thomson Ave., Cambridge CB3 0HE, UK.
   \and Kavli Institute for Cosmology, University of Cambridge, Madingley Road, Cambridge CB3 0HA, UK.
   \and Department of Physics \& Astronomy, University College London, Gower Street, London WC1E 6BT, UK.
   \and Centro de Astrobiolog\'ia (CAB), CSIC--INTA, Cra. de Ajalvir Km.~4, 28850 -- Torrej\'on de Ardoz, Madrid, Spain.
   \and Physics Department, Tor Vergata University of Rome, Via della Ricerca Scientifica 1, 00133 Rome, Italy%
   \and INFN - Rome Tor Vergata, Via della Ricerca Scientifica 1, 00133 Rome, Italy.
   \and Institut d’Astrophysique de Paris, Sorbonne Université, CNRS,  UMR 7095, 98 bis bd Arago, 75014 Paris, France.%
   \and Dipartimento di Fisica e Astronomia ‘Augusto Righi’, Università degli Studi di Bologna, via P. Gobetti, 93/2, 40129 Bologna, Italy.%
   \and Smithsonian Astrophysical Observatory, 60 Garden Street MS09, Cambridge, MA 02138, USA.
   \and Observatoire de Paris, LUX, CNRS, PSL University, 75014, Paris, France.
   \and NASA Goddard Space Flight Center, Code 662, Greenbelt, MD 20771, USA%
   \and INFN – National Institute for Nuclear Physics, via Valerio 2, I-34127 Trieste, Italy.
   \and Department of Physics, University of Napoli “Federico II”, via Cinthia 9, 80126 Napoli, Italy.
   \and INAF - Osservatorio Astronomico di Capodimonte, via Moiariello 16, 80131 Napoli, Italy.
   \and Centre for Extragalactic Astronomy, Department of Physics, Durham University, South Road, Durham DH1 3LE, UK.
   \and Center for Astrophysics - Harvard \& Smithsonian, Cambridge, MA 02138, USA.
   \and Steward Observatory, University of Arizona, Tucson, Arizona, USA.
   \and INAF - Osservatorio Astronomico di Padova, vicolo dell'Osservatorio 5, I-35122, Padova, Italy.
   \and European Space Agency, ESTEC, Keplerlaan 1, 2201 AZ Noordwijk, The Netherlands.
   \and DiSAT, Università degli Studi dell’Insubria, via Valleggio 11, I-22100 Como, Italy.
   \and INFN - Sezione di Milano-Bicocca, piazza della Scienza 3, I-20126 Milano, Italy.
   \and INAF - Osservatorio Astronomico di Brera, via E. Bianchi 46, I-23807 Merate, Italy.
   \and ASI - Agenzia Spaziale Italiana, Via del Politecnico snc, I-00133 Roma, Italy.
   \and Dipartimento di Fisica, Università di Roma La Sapienza, Piazzale Aldo Moro 2, I-00185 Roma, Italy.
   \and INFN, Sezione Roma1, Dipartimento di Fisica, Università di Roma La Sapienza, Piazzale Aldo Moro 2, I-00185 Roma, Italy.
   \and Sapienza School for Advanced Studies, Viale Regina Elena 291, I- 00161 Roma, Italy.
   \and INAF - Istituto di Astrofisica Spaziale e Fisica Cosmica Milano, Via A. Corti 12, 20133 Milano, Italy.
   }
   \date{Received XX, 2026; accepted YY, 2026}

\abstract
{Luminous quasars (QSOs) at the Epoch of Reionization (EoR, $z\gtrsim6$) host billion-solar-mass black holes assembled within the first Gyr of cosmic time. Their X-ray emission probes the innermost accreting regions and provides direct constraints on the physical state of the X-ray emitting corona during early supermassive black hole (SMBH) growth. Previous results from the HYPerluminous quasars at the Epoch of ReionizatION (HYPERION) sample of $z \gtrsim 6$ QSOs powered by the most rapidly accreted SMBHs suggested a link between X-ray spectral slope and long-term growth efficiency. We aim to test whether the systematically steep X-ray spectra observed in luminous $z \gtrsim 6$ QSOs are a general property of early SMBH growth or instead are a peculiarity of the most efficiently grown SMBHs; to this end, we investigate this property in an expanded QSO sample, deliberately extended to include mainly the lowest SMBH mass growth efficiencies. We analysed the largest and most X-ray sensitive sample of 28 luminous QSOs at $z \simeq 5.85-7.64$ (after excluding two sources with insufficient photon statistics the correlation analysis is based on 26 QSOs), combining the original HYPERION targets, luminous non-HYPERION sources, and newly observed QSOs, powered by moderately accreting SMBHs, from our recent \xmm\ Large Programme. We re-derived the black hole masses and growth indicators for the entire sample using a homogeneous methodology, incorporating the latest JWST-based spectroscopic measurements to ensure the most consistent and high-quality characterisation across all targets. We found no statistically significant correlation between coronal spectral parameters and the BH growth indicators, while luminous QSOs at $z \gtrsim 6$ exhibit systematically steep X-ray continua across the whole sampled range, characterised by high photon indices and/or low high-energy cutoffs. This property is consistent with cold coronae, possibly associated with rapidly accreting regimes. Finally, we show that if the steep X-ray continuum slopes are a common feature of the early Universe AGN population, they may introduce a significant observational bias, reducing the observed count rates at fixed intrinsic luminosity, in future extragalactic surveys targeting the EoR Universe, with consequences for the detectability of steep-spectrum X-ray sources and therefore for our ability to conduct an unbiased census of the early AGN population.}

\maketitle
\nolinenumbers
\section{Introduction}
\label{sect:intro}
The detection of highly luminous quasars (QSOs) at the Epoch of Reionization (EoR, $z \gtrsim 6$), powered by supermassive black holes (SMBHs) with masses reaching $M_{\rm BH}\sim10^{9-10}\,M_\odot$, represents one of the most stringent tests for theoretical models of BH formation and early growth \citep[e.g.][]{2010A&ARv..18..279V,2020ARA&A..58...27I,2023ARA&A..61..373F}. If the progenitors of these SMBHs formed at $z\sim20$--30, the rapid build-up necessary to assemble their masses within the first $\lesssim1$\,Gyr of cosmic time requires either already massive BH seeds ($\gtrsim10^{3}$--$10^{5}\,M_\odot$), episodes of super-Eddington accretion, and/or a combination of both \citep{Trinca2022}. Distinguishing between these possibilities remains a central unresolved problem in modern astrophysics.\\
\indent Over the last two decades, extensive near-infrared and sub-millimetre campaigns have yielded increasingly precise constraints on the rest-frame UV and optical emission, as well as on the host-galaxy characteristics, of broad-line, radio-quiet, $z \gtrsim 6$ QSOs \citep[e.g.][]{2018ApJ...854...97D,Shen2019,Wang_2021,2017ApJ...849...91M,2023MNRAS.523.1399D,2023ApJ...946L..45T,2024A&A...689A.220T,Saccheo2025,2025A&A...695A..23S}, showing that their observed properties are remarkably similar to those of luminous QSOs at lower redshift.\\
\indent In comparison, X-ray investigations of these objects have been limited by poor photon statistics, frequently limiting studies to mostly marginal detections or limited stacked analyses \citep[e.g.][]{Vignali03,2017A&A...603A.128N,Vito2019,2019A&A...631A.120S,2022ApJ...924L..25Y}. This shortcoming is especially important because X-ray radiation carries information on the innermost regions of the accretion flow and thus provides an almost instantaneous probe of the disc–corona system, which is tightly coupled to SMBH growth. Indeed, the X-ray emission of AGN and QSOs is due to high-energy processes that occur in the hot corona, which contains high-energy electrons and is located near the SMBH \citep{1991ApJ...380L..51H,1994ApJ...432L..95H,1997ApJ...476..620H,2003MNRAS.341.1051M,2009Natur.459..540F,2012MNRAS.422..129Z,2013MNRAS.431.2441D}. The hot electrons interact with thermal UV/optical photons emitted by the accretion disc through inverse-Compton scattering, which boosts the energy of the photons into the X-ray band. This process produces a broad power-law continuum in the X-ray spectrum \citep[e.g.][]{1980A&A....86..121S,1993ApJ...413..507H} with photon indices $\Gamma\sim1.8-2.0$ \citep{Dadina08,Piconcelli2005} and a cutoff at high energy \citep[e.g.][]{Fabian2015,Ricci_2018,2018A&A...614A..37T}.\\
\indent To overcome the above-mentioned limitations, the HYPerluminous quasars at the Epoch of ReionizatION (HYPERION) programme was initiated. The HYPERION sample consists of luminous ($L_{\rm bol}\gtrsim10^{47}$\,erg\,s$^{-1}$) QSOs known by 2020, powered by SMBHs that appear to have experienced the most rapid growth among sources at comparable epochs.\\
\indent In particular, the HYPERION sample has been selected to include QSOs powered by SMBHs characterised by highly efficient growth tracks. We parameterise this long-term efficiency via the \mseed\ parameter, defined as the initial mass required for a hypothetical $z = 20$ seed BH undergoing continuous, exponential accretion at the Eddington rate ($L_{\rm bol}/L_{\rm Edd}=\lambda_{\rm Edd} =1 $) to assemble the final SMBH. For the HYPERION targets this idealised track, which we label using this same \mseed\ parameter, requires a starting value of \mseed $>1000 M_{\odot}$  (\citealp{Zappacosta2023}, hereafter \citetalias{Zappacosta2023}). The \mseed\ parameter can be regarded as a proxy for the SMBH growth efficiency integrated over the SMBH assembly period, where higher \mseed\ values correspond to a more rapid and efficient growth history (for further details see \citealp{Tortosa2024}, hereafter {\citetalias{Tortosa2024}). A \xmm\ Multi--Year Heritage programme (X-HP), comprising 2.4~Ms of observations dedicated to the HYPERION sample, provided the most sensitive and statistically sound X-ray characterisation of the QSOs at EoR powered by SMBHs that experienced the most extreme and challenging growth at $z \gtrsim 6$ \citepalias{Zappacosta2023}.\\
\indent Results from the X-HP revealed that the X-ray spectra of $z \gtrsim 6$ QSOs are on average significantly steeper ($\langle\Gamma\rangle=2.4\pm0.1$) than those of comparable lower-$z$ sources \citepalias{Zappacosta2023,Tortosa2024}. This behaviour is consistent with intrinsically colder coronae. A marginal correlation has been reported between $\Gamma$ and \mseed\, \citepalias{Tortosa2024} supporting the hypothesis that X-ray spectral characteristics can reflect the growth history of SMBHs. Moreover, \citetalias{Tortosa2024} reports a significant correlation between $\Gamma$ and the velocity shift of the \ion{C}{iv} emission line in QSOs at $z \gtrsim 6$, pointing to a physical connection between accretion disc winds and coronal emission.\\
\indent However, a critical question remains unresolved: are the systematically steep X-ray continua observed in $z \gtrsim 6$ QSOs a universal feature of early SMBH growth, or are they confined to the $L_{\rm bol}-M_{\rm BH}$ parameter space probed by the HYPERION sample? To address this question, a new \xmm\ Large Programme (X-LP n.094377, PI L. Zappacosta) to observe 5 targets for a total of 600 ks, has been recently approved. The X-LP extends the HYPERION dataset to QSOs with either smaller and larger \mseed\ than originally probed, but with comparable $L_{\rm bol}$. This will allow a more robust exploration of the link between X-ray properties, disc winds, and SMBH growth indicators, in combination with HYPERION and archival and proprietary high-quality X-ray observations of non-HYPERION QSOs.\\
\indent In this paper, we exploit a sample of 28 $z \gtrsim 6$ QSOs, including those from the new X-LP. The paper is organised as follows. In Sect.~\ref{sect:sample_and_data_red}, we present the sample. In Sect.~\ref{sect:Analysis} we present the spectral and correlation analysis performed in this work and their results. In Sect.~\ref{sect:Discussion}, we discuss the X-ray properties of the $z \gtrsim 6$ QSO population. Finally, our conclusions are summarised in Sect.~\ref{sect:conclusions}.\\
\indent Standard cosmological parameters (H$_{0}$=70\,km\,s$^{-1}$\,Mpc$^{-1}$, $\Omega_{\Lambda}$=0.73 and $\Omega_m$=0.27) are adopted throughout the paper. Errors are reported at the 68\% confidence level with upper and/or lower limits quoted at the 90\% confidence level, if not stated otherwise.
\section{Sample selection and data presentation}
\label{sect:sample_and_data_red}
\subsection{The sample}
\label{subsec:sample}
\begin{figure}
\centering
\includegraphics[width=0.9\columnwidth]{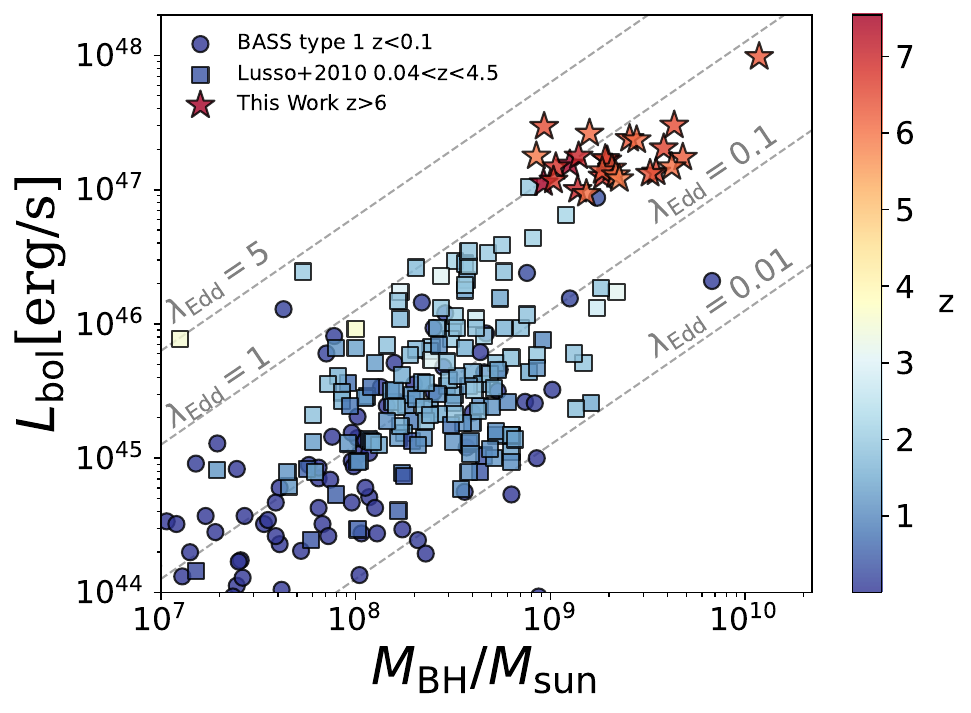}
\caption{Distribution of the sample in the $M_{\rm BH}$--$L_{\rm bol}$ plane. Symbols are colour-coded by redshift. We compared our sample ($M_{\rm BH}$ values plotted here are, for simplicity, the mean values derived using all the available values reported in Tab.~\ref{table:masses}) with the X-ray selected type 1 AGN in XMM-COSMOS $0.04<z<4.5$ from \citealt{Lusso2010} and the values of the X-ray unobscured ($N_H < 10^{21}\,\rm cm^{-2}$) type~I $z<0.1$ AGN from the BASS sample from \citealt {2022ApJS..261....5M}.}
\label{fig:MBH_Lbol}
\end{figure}
The analysis presented in this work is based on a large sample of 28 broad-line, unlensed, radio-quiet, luminous ($L_{\rm bol} \gtrsim 10^{47}\,\mathrm{erg\,s^{-1}}$) QSOs at $z \gtrsim 6$. The sample is defined by the availability of (i) SMBH mass estimates from rest-frame UV/optical spectroscopy and (ii) X-ray observations with \xmm\ or \textit{Chandra}, either from our dedicated programmes (X-HP and X-LP, for the HYPERION and X-LP targets, selected independently of their X-ray properties) or from archival data of sufficient quality for spectral analysis. For the spectral, statistical and correlation analyses we further require $>30$ net counts in the 0.3--7~keV energy band; 26 of the 28 QSOs satisfy this criterion, while J0313-1806 and PSO~J183-12 do not and are treated through upper limits (see Sect.~\ref{subsect:source_detection}).
The distribution of the sample in the $M_{\rm BH}$-$L_{\rm bol}$ plane is shown in Fig.~\ref{fig:MBH_Lbol}.
We considered:
\begin{itemize}
    \item The full HYPERION sample composed by 18 QSOs, whose observations mostly belong to the dedicated X-HP presented in \citetalias{Zappacosta2023} and further analysed in \citetalias{Tortosa2024}. This sample includes the QSOs whose SMBHs span the upper end of the SMBH growth-rate distribution (i.e., $\log($\mseed$/M_{\odot})>3$) at EoR;
    \item Five non-HYPERION QSOs covering the range $2.5\lesssim\log($\mseed$/M_{\odot})\lesssim3.3$;
    \item Five newly observed QSOs from the recently approved X-LP, extending the \mseed\ parameter space to $2.1\lesssim\log($\mseed$/M_{\odot})\lesssim2.4$ and $\log($\mseed$/M_{\odot})>4$ with four and one QSOs, respectively. These QSOs have \mseed\ chosen to be representative of the central 90\% of the distribution of $M_{\rm BH}$ and $L_{\rm bol}$ in the HYPERION and non HYPERION samples considered in \citetalias{Zappacosta2023} and \citetalias{Tortosa2024}. They were selected among the known population of optically and near-infrared identified luminous QSOs, with reliable SMBH mass estimates (from Mg~\textsc{ii} single epoch virial estimator) available in the literature, and were chosen to uniformly populate both the low and high ends of the \mseed\ distribution of the whole sample. 
\end{itemize}
The \mseed\ values used for the sample selection were derived from Mg~\textsc{ii}-based masses (from \citetalias{Tortosa2024} for the HYPERION and non-HYPERION QSOs, and from the literature available at the time of the observing proposal for the X-LP targets). These differ from the updated values re-derived in this work (Tab.~\ref{table:masses}), which are used throughout the analysis.
\begin{table*}
\caption{Sample of $z \gtrsim 6$ QSOs considered in this work with best-fit parameters from the X-ray spectral analysis. $M_{\rm BH}$, and \mseed, are shown in Tab.~\ref{table:masses}. }
\label{table:sources}
\centering
\small
\begin{tabular}{lccccccccc}
\hline\hline
Target &  $z^{\rm a}$& $\log(L_{\rm bol})^{\rm b}$ & $\log(\lambda L_{\rm 2500\,\AA})^{\rm b}$ &  $\Gamma$ & $E_{\rm c}^{\rm c}$ &log($L_{2-10\,\rm keV}^{\rm d}$) & $\alpha_{\rm OX}$ & $\Delta\alpha_{\rm OX}^{\rm e}$ & $\log(\kappa_{\rm bol}^{\rm X})$\\[0.1cm]
       &     & $[\rm erg\,s^{-1}]$ &$[\rm erg\,s^{-1}]$  &   & [keV]  &$[\rm erg\,s^{-1}]$ &  &  &   \\[0.1cm]
\hline
\multicolumn{10} {c}{HYPERION QSOs} \\[0.1cm]
ULAS~J1342+0928 & 7.541 &$47.05\pm 0.04$ &$46.60\pm0.02$& $2.93^{+0.43}_{-0.37}$ & $8_{-5}^{+40}$  & $45.22\pm0.08$ & $-1.49\pm 0.06$ & $+0.18$ &$1.83 \pm 0.39$ \\[0.1cm]
J1007+2115      & 7.494 & $47.19\pm0.07$ &$46.75\pm0.03$ & $2.66^{+0.78}_{-0.78}$ & $14.8_{-10}^{+15}$& $44.50\pm0.09$ & $-1.86\pm0.08$ & $-0.17$ & $2.69 \pm  0.56$\\[0.1cm]
ULAS~J1120+0641 & 7.087 &$47.22\pm0.05$ &$46.84\pm0.05$& $2.65^{+0.35}_{-0.32}$ & $7.5_{-4}^{+20}$  & $45.39\pm0.06$ & $-1.52\pm0.07$ & $+0.19$ &$1.83 \pm 0.38$ \\[0.1cm]
DELS~J0038-1527 & 7.021 &$47.25\pm0.06$ &$46.87\pm0.03$ & $3.25^{+1.4}_{-1.3}$ &$>3.5$ & $45.01\pm0.10$ & $-1.89\pm0.08$ & $-0.18$ & $2.24 \pm 0.48$ \\[0.1cm]
DES~J0252-0503  & 6.990 & $47.00\pm0.06$ &$46.59 \pm 0.04$& $2.63^{+0.92}_{-0.92}$ & $3.9_{-2}^{+30}$  & $45.08\pm0.05$ & $-1.59\pm0.13$ & $+0.08$ & $1.92 \pm 0.41 $ \\[0.1cm]
VDES~J0020-3653 & 6.834 &$47.10\pm0.06$ &$46.69 \pm 0.05$ & $2.73^{+0.55}_{-0.52}$ & $9.2_{-5}^{+25}$  & $45.12\pm0.08$ & $ -1.56\pm0.02$ & $+0.13$ & $1.98 \pm 0.41$ \\[0.1cm]
VHS~J0411-0907  & 6.824 &$47.17\pm0.04$ &$46.74\pm0.02$ & $1.84^{+0.48}_{-0.48}$ & $>50$  & $45.13\pm0.05$ & $-1.83\pm0.08$ & $-0.14$ & $2.04 \pm 0.44$\\[0.1cm]
VDES~J0244-5008 & 6.724 &$47.07\pm0.04$ &$46.58\pm0.03$& $2.35^{+0.30}_{-0.25}$ &$>25$  & $45.28\pm0.04$ & $-1.54\pm0.02$ & $+0.13$ & $1.79 \pm 0.38$ \\[0.1cm]
PSO~J231.6-20.8 & 6.587 &$47.31\pm0.05$ &$46.70\pm0.04$ & $2.30^{+0.62}_{-0.67}$ & $>13$ & $45.09\pm0.09$ & $-1.90\pm0.16$ & $-0.21$ & $2.22 \pm 0.44$ \\[0.1cm]
PSO~J036.5+03.0 & 6.533 &$47.13\pm0.04$ &$46.79\pm0.03$& $3.05^{+1.1}_{-0.85}$ &$10.5_{-5}^{+50}$  & $44.95\pm0.11$ & $-1.67\pm0.05$ & $+0.03$ & $2.18 \pm 0.47$\\[0.1cm]
VDES~J0224-4711 & 6.526 &$47.23\pm0.05$ &$47.00\pm0.02$ & $2.08^{+0.20}_{-0.20}$ &$>28$ & $45.58\pm0.05$ & $-1.73\pm0.02$ & $-0.00$ &$1.65 \pm  0.39$\\[0.1cm]
PSO~J011+09  & 6.444 &$47.47\pm0.04$ &$46.39\pm0.02$ & $2.52^{+0.70}_{-0.70}$ &$>13$ & $45.04\pm0.11$ & $-1.76\pm0.07$ & $-0.12$ & $2.43 \pm 0.42$\\[0.1cm]
SDSS~J1148+5251 & 6.422 &$47.48\pm0.04$ &$46.95\pm0.02$& $2.38^{+0.39}_{-0.39}$ &$16.4_{-8}^{+60}$  & $45.18\pm0.08$ & $-1.79\pm0.05$ & $-0.07$ & $2.30 \pm 0.48$\\[0.1cm]
PSO~J083.8+11.8 & 6.346 & $47.16\pm0.02$&$46.71\pm0.03$& $1.89^{+0.39}_{-0.37}$ &$>20$   & $45.02\pm0.08$ & $-1.64\pm0.01$ & $+0.05$ & $2.14 \pm 0.48$ \\[0.1cm]
SDSS~J0100+2802 & 6.300 & $47.99\pm0.03$ & $47.54\pm0.03$ & $2.41^{+0.15}_{-0.09}$ &$31^{+53}_{-13}$& $45.76\pm0.03$ & $-1.72\pm0.01$ & $+0.09$ & $2.23 \pm 0.49$\\[0.1cm]
ATLAS~J025-33   & 6.294 &$47.38\pm0.04$ & $46.93\pm0.01$& $2.03^{+0.42}_{-0.37}$ &$>50$ & $45.24\pm0.03$ & $-1.83\pm0.02$ & $-0.11$ & $2.14 \pm 0.43$ \\[0.1cm]
CFHQS~J0050+3445& 6.246 & $47.24\pm0.06$& $46.84\pm0.02$ & $2.02^{+0.53}_{-0.55}$ &$34^{+13}_{-10}$ & $45.07\pm0.08$ & $-1.71\pm0.04$ & $-0.00$ & $2.17 \pm 0.44$\\[0.1cm]
ATLAS~J029-36   & 6.027 & $47.17\pm0.06$ &$46.66\pm0.03$& $2.54^{+0.24}_{-0.21}$ &$>23$ & $44.93\pm0.08$ & $-1.67\pm0.03$ & $+0.01$ & $2.24 \pm 0.49 $\\[0.1cm]
\hline
\multicolumn{10} {c}{Non-HYPERION QSOs} \\[0.1cm]
PSO~J159-02     & 6.380$^{\rm f}$ &$47.12\pm0.04^{\rm h}$ & $46.59\pm0.05^{\rm h}$ & $1.86^{+0.45}_{-0.40}$ & $>23$ & $45.41\pm0.11$ & $-1.66\pm0.02$ & $+0.01$ & $1.71 \pm 0.38$ \\[0.1cm]
SDSS~J1030+0524 & 6.308 &$47.14\pm0.14$ & $46.66\pm0.03$& $2.19^{+0.20}_{-0.19}$ & $29^{+58}_{-13}$  & $45.59\pm0.09$ & $-1.58\pm0.04$ & $+0.10$ & $1.55 \pm 0.31$ \\[0.1cm]
SDSS~J1602+4228 & 6.090$^{\rm g}$&$47.42\pm0.04^{\rm h}$ & $46.91\pm0.05^{\rm h}$& $1.97^{+0.73}_{-0.74}$ & $>12$  & $45.53\pm0.16$ & $-1.52\pm0.03$ & $+0.20$ & $ 1.89 \pm 0.30 $\\[0.1cm]
PSO~J308-21     & 6.234$^{\rm f}$&$47.37\pm0.01$ & $46.58\pm0.03$  & $2.45^{+0.38}_{-0.41}$ & $>9.8$ & $45.36\pm0.08$ & $-1.57\pm0.03$ & $+0.10$ & $2.01\pm0.40$ \\[0.1cm]
SDSS~J1306+0356 & 6.034$^{\rm f}$& $47.11\pm0.04$ &$46.63\pm0.02$  & $1.85^{+0.35}_{-0.26}$ & $>35$  & $45.23\pm0.05$ & $-1.41\pm0.02$ & $+0.27$ & $1.88 \pm 0.38$\\[0.1cm]
\hline
\multicolumn{10} {c}{X-LP} \\[0.1cm]
J0313-1806       & 7.640$^{\rm f}$ & $47.07\pm0.07^{\rm h}$ & $46.69\pm0.09^{\rm h}$ & $2.0^{\rm i}$                 &              & $<43.97^{\rm j}$ & $<-2.07^{\rm j}$ & $<-0.38^{\rm j}$ & $>3.16^{\rm j}$\\[0.1cm]
    &    &    &     &  $2.4^{\rm i}$                 &              & $<43.90^{\rm j}$ & $<-2.04^{\rm j}$ & $<-0.35^{\rm j}$ & $>3.09^{\rm j}$\\[0.1cm]
SDSS~J2054-0005  & 6.039$^{\rm f}$ &$46.97\pm0.06^{\rm h}$ & $46.35\pm0.18^{\rm h}$ & $2.86^{+0.47}_{-0.51}$ & $>6.1$            & $44.46\pm0.12$ & $-1.61 \pm 0.05$ & $+0.03$ & $2.51 \pm 0.21$\\[0.1cm]
SDSS~J1411+1217  & 5.920 &$47.25\pm0.04^{\rm h}$  & $46.72\pm0.05^{\rm h}$  & $2.58^{+0.28}_{-0.29}$ & $8.3_{-5}^{+19}$  & $44.84\pm0.15$ & $-1.80 \pm 0.08$ & $-0.11$ & $2.41 \pm 0.65$ \\[0.1cm]
PSO~J183-12  &  5.860 & $47.27\pm0.04$ & $46.67\pm0.04$ & $2.0^i$  &  & $<44.72^{\rm j}$ & $<-1.83^{\rm j}$ & $<-0.15^{\rm j}$ & $>2.55^{\rm j}$\\[0.1cm]
    &    &    &     & $2.4^{\rm i}$ &  & $<44.42^{\rm j}$  & $<-1.86^{\rm j}$ & $<-0.18^{\rm j}$ & $>2.85^{\rm j}$\\[0.1cm]
PSO~J025-11  & 5.850$^{\rm f}$ & $47.08\pm0.05$ & $46.56\pm0.02$& $2.45^{+0.68}_{-0.65}$ & $>9.8$  & $44.10\pm0.08$ & $-2.02 \pm 0.07$ & $-0.35$ & $2.98 \pm 0.97$ \\[0.1cm]
\hline\hline
\end{tabular}
\tablefoot{a: estimated from the Mg\,\textsc{ii} emission line if not stated otherwise; b: from \citealt{Saccheo2025} if not stated otherwise; c: obtained from a power-law with an exponential cutoff energy (\texttt{zcutoffpl} model in \texttt{xspec}) with $\Gamma$ fixed to a value of $1.9$; d: $L_{2-10\,\rm keV}$ is measured in the 2--10\,keV rest-frame energy band; e: $\Delta\alpha_{\rm OX} = \alpha_{\rm OX}({\rm observed}) - \alpha_{\rm OX,LR16}(L_{2500\,\AA})$, computed from the \citealt{Lusso2016} relation;  f: $z$ estimated from the [C\,\textsc{ii}] emission line; g: $z$ estimated from the Ly\,$\alpha$ emission line; h: $L_{2500\,\AA}$ and $L_{\rm bol}$ derived from the rest-frame UV magnitudes applying spectral templates from \citealt{Saccheo2025}; i: fixed parameter; j: 90\% upper limit from EPIC-PN photometry from Tab.~\ref{table:counts} assuming a power-law with the fixed $\Gamma$.\\}
\end{table*}
We report in Tab.~\ref{table:sources} the list of all the QSOs considered in this work and their general properties. To ensure a homogeneous determination of the BH mass ($M_{\rm BH}$) and \mseed\ across the full sample, we re-derived all virial mass estimates using consistent scaling relations and cosmological assumptions (see App.~\ref{app:masses} for more details). The monochromatic ultraviolet luminosity at 2500~\AA\ ($L_{2500\,\AA}$) and the bolometric luminosity ($L_{\rm bol}$) were primarily taken from \citealt{Saccheo2025} based on a homogeneous X-ray-to-optical spectral energy distribution modelling, ensuring consistency across the sample. For the few sources not considered by \citealt{Saccheo2025}, i.e. SDSS~J1602+4228, PSO~J159-02, SDSS~J1411+1217, SDSS~J2054-0005, J0313-1806, we estimated $L_{2500\,\AA}$ and $L_{\rm bol}$ starting from the rest-frame using the $y$, $J$, $H$, $K$, \textit{WISE}~$W1$, and \textit{WISE}~$W2$ magnitudes reported by \citealt{RossCross2020} \citep[except J0313-1806 for which we used the values from][]{Wang_2021}. In these cases, the magnitudes were converted into monochromatic luminosities and we extrapolated $L_{2500\,\AA}$ and $L_{\rm bol}$ by adopting the same quasar spectral templates described in \citealt{Saccheo2025}. This approach guarantees that all luminosity estimates used in this work are consistent and minimises systematic differences arising from heterogeneous assumptions on the spectral shape.
\subsection{Data reduction and source detection}
\label{subsect:source_detection}
\begin{table}
\caption{Targets net counts from fixed aperture photometry on the \xmm\ EPIC detectors and on the \textit{Chandra} ACIS-S detector in the 0.3--7\,keV energy band.}
\label{table:counts}  
\centering 
\small
\begin{tabular}{llll}   
\hline\hline                
Target &  PN & MOS1 & MOS2 \\[0.1cm]
\hline
ULAS~J1342+0928	& $108.0_{-28.0}^{+27.0}$ & $66.3_{-17.9}^{+18.4}$ &$<59.5$\\[0.1cm]
J1007+2115 +2115      & $59.8_{-21.0}^{+21.4}$ & $<59.1$ &$<28.3$ \\[0.1cm]
ULAS~J1120+0641 & $78.4_{-15.0}^{+15.4}$ & $70.9_{-13.9}^{+14.4}$ &$42.7_{-13.8}^{+14.3}$\\[0.1cm]
DELS~J0038-1527 & $66.0_{-27.0}^{+28.0}$ & $<49.7$ &$<32.1$\\[0.1cm]
DES~J0252-0503  & $79.0_{-24.0}^{+26.0}$ & $<73.0$  &$<53.0$\\[0.1cm]
VDES~J0020-3653 & $29.2_{-11.0}^{+11.6}$ & $20.81_{-8.5}^{+9.1}$ &$<24.8$\\[0.1cm]
VHS~J0411-0907  & $92.7_{-20.0}^{+20.4}$ & $56.9_{-18.4}^{+18.9}$ &$54.0_{-15.9}^{+16.5}$\\[0.1cm]
VDES~J0244-5008 &$98.7_{-15.4}^{+15.9}$  & $36.46_{-10.1}^{+10.6}$ &$31.61_{-10.3}^{+10.8}$\\[0.1cm]
PSO~J231.6-20.8 &$98.5_{-33.6}^{+34.0}$  &$<54.4$ & $<50.0$ \\[0.1cm]
PSO~J036.5+03.0 &$78.4_{-21.4}^{+21.8}$ & $<37.8$ &$<28.7$\\[0.1cm]
VDES~J0224-4711 &$84.5_{-12.7}^{+13.1}$ & $35.9_{-12.5}^{+12.8}$ &$30.53_{-11.0}^{+11.4}$\\[0.1cm]
PSO~J011+09     &$64.6_{-25.4}^{+25.7}$ & $<45.2$ &$<36.2$\\[0.1cm]
SDSS~J1148+5251 &$89.8_{-19.9}^{+20.3}$ & $<44.5$ &$<44.0$\\[0.1cm]
PSO~J083.8+11.8 &$50.0_{-17.1}^{+17.5}$ & $44.6_{-13.8}^{+14.2}$ &$32.3_{-13.1}^{+13.5}$\\[0.1cm]
SDSS~J0100+2802 &$240.7_{-21.4}^{+21.9}$ & $81.0_{-12.7}^{+13.1}$ &$52.7_{-11.0}^{+11.4}$\\[0.1cm]
ATLAS~J025-33   &$84.2_{-17.4}^{+17.8}$ & $75.5_{-16.9}^{+17.3}$ &$74.7_{-15.6}^{+16.0}$\\[0.1cm]
CFHQS~J0050+3445 &$34.1_{-13.8}^{+14.3}$ & $<44.5$ &$<58.7$\\[0.1cm]
ATLAS~J029-36   &$203.0_{-32.0}^{+32.0}$ & $<71.0$ &$<65.0$\\[0.1cm]
PSO~J159-02     & $54.1_{-11.5}^{+12.0}$&$22.76_{-7.03}^{+7.54}$&$18.4_{-6.46}^{+6.95}$\\[0.1cm]
SDSS~J1030+0524 & $165.2_{-22.6}^{+23.0}$&$75.3_{-12.3}^{+12.8}$&$107.5_{-13.4}^{+13.9}$\\[0.1cm]
J0313-1806      &$18.2^{+6.5}_{-6.1}$ & $<13.54$ &$<9.85$\\[0.1cm]
SDSS~J2054-0005 &$43.7_{-15.8}^{+16.1}$ & $<33.6$ &$21.2_{-9.0}^{+9.3}$\\[0.1cm]
SDSS~J1411+1217 &$62.6_{-15.8}^{+16.2}$ & $<30.01$ &$<27.92$\\[0.1cm]
PSO~J183-12     &$<10.59$ & $<16.92$ & $<15.21$ \\[0.1cm]
PSO~J025-11     &$44.3_{-15.6}^{+16.1}$ & $<35.0$ &$<23.49$\\[0.1cm]
\hline
\hline
Target &  \multicolumn{3} {c}{ACIS-S} \\[0.1cm]
\hline
SDSS~J1602+4228 & \multicolumn{3} {c}{$38.6_{-6.5}^{+6.9}$}\\[0.1cm]
PSO~J308-21     & \multicolumn{3} {c}{$85.1_{-8.9}^{+7.5}$}\\[0.1cm]
SDSS~J1306+0356 & \multicolumn{3} {c}{$145.2_{-6.1}^{+6.0}$} \\[0.1cm]
\hline\hline 
\end{tabular}
\end{table}
The data reduction procedure is described in App.~\ref{app:data_red}. In App.~\ref{app:xmm_data_red}, we provide the details of the new \xmm\ observations that were processed and analysed for this study, namely those from the recent X-LP, along with the X-HP observations that were not included in \citetalias{Tortosa2024}. This set includes PSO~J083.8+11.8, which had already been analysed in \citetalias{Zappacosta2023} but was excluded from \citetalias{Tortosa2024} due to the lack of an available measurement of the velocity of the outflow coming from the C\,\textsc{iv} line, $v_{\rm C\,\textsc{iv}}$.\\
\indent Several of the non-HYPERION QSOs discussed in \citetalias{Tortosa2024} were observed with \textit{Chandra}. In this work, we re-analysed these \textit{Chandra} data. The corresponding data reduction procedure is described in detail in App.~\ref{app:Chandra_data_red}.\\
\indent Source detection significance was quantified using the no-source binomial probability, following the methodology described in \citealt{2007ApJ...657.1026W} and adopted by \citealt{Vito2019}. Source and background counts were extracted from circular source regions and surrounding background regions in the combined 0.3--7~keV images of the three EPIC cameras or of ACIS-S (for PSO~J308-21, the individual \textit{Chandra} OBSIDs were reprojected and merged with \texttt{reproject\_obs} and \texttt{merge\_obs}). Sources with a no-source binomial probability below the value corresponding to a 99\% confidence level were considered detected.\\
\indent For significantly detected sources, net counts in the 0.3--7~keV band were estimated by subtracting the expected number of background counts from the total counts in the source region. Count uncertainties were estimated assuming Poisson statistics for both the source and background counts, and propagated following the same approach adopted in \citealt{Vito2019}. The net counts for each target from fixed aperture photometry on the EPIC and ACIS-S detectors in the 0.3--7~keV energy band are reported in Tab.~\ref{table:counts}.\\
\indent All targets were detected except J183-12. J0313-1806 is formally detected in EPIC-PN with $< 20$ counts, and does not reach the minimum threshold of $\sim 30$ net counts in the 0.3--7~keV band required for a reliable spectral analysis, following the criterion adopted in \citetalias{Tortosa2024}. Moreover, the source extraction region is affected by the presence of two nearby X-ray sources (see App.~\ref{app:xmm_data_red}), whose PSF wings contaminate the QSO emission, making the measurement of its flux and therefore its detection particularly uncertain. For these reasons, J183-12 and J0313-1806 are not included in the subsequent spectral, statistical and correlation analyses.
\section{Analysis}
\label{sect:Analysis}
\subsection{Spectral analysis}
\label{sect:spectral_analysis}
The spectral analysis was performed using the \texttt{xspec} v.12.12.1 software package \citep{Arnaud1996}. Following \citetalias{Zappacosta2023} and \citetalias{Tortosa2024}, we performed the modelling using the Cash statistic with a Poisson likelihood for the background spectrum \citep[W-stat in \texttt{xspec},][]{1979ApJ...228..939C,1979ApJ...230..274W}, rather than a direct background subtraction. We adopted a simple power-law model, modified by Galactic absorption. All sources are Type~1 QSOs with very low levels of intrinsic absorption (on average $<10^{21}\rm cm^{-2}$; \citetalias{Zappacosta2023}) which, given the high-$z$ of the sources, have only a negligible impact on the continuum slope, affecting only the softest observed energies (0.3~keV, $\sim2$~keV rest-frame). We further included a constant in the model to account for possible flux variations between observations obtained at different epochs (parametrised by \texttt{const}*\texttt{tbabs}*\texttt{zpowerlaw} in \texttt{xspec}). For the objects with multi-epoch observations, no flux or spectral variability has been found, thus we fixed the constant to $1$ in the following analysis. For the Galactic column density at the position of the sources, we adopted the \citealp{HI4PI2016} maps. The analysis was performed over the optimised 0.3--7\,keV energy range (see App. C in \citetalias{Zappacosta2023}) which corresponds to a rest-frame $\sim2-50$\,keV energy band. We jointly modelled the EPIC-PN and MOS spectra, linking $\Gamma$ and leaving it free to vary together with the normalisation.\\
\indent The analysis of the \xmm\ observations of the HYPERION sample available up to June 2024, together with that of additional non-HYPERION QSOs with archival, good--quality X-ray data, was first presented in \citetalias{Tortosa2024}. In this work, we adopt the spectral results reported by \citetalias{Tortosa2024} for the \xmm\ sources, while we have reprocessed and re-analysed the \textit{Chandra} data and the new \xmm\ observations. The parameters derived from our spectral analysis are listed in Tab.~\ref{table:sources}, alongside those reported in \citetalias{Tortosa2024}.\\
\indent We additionally derived the intrinsic (Galactic absorption corrected) rest-frame 2–10 keV luminosity ($L_{\mathrm{2-10\,keV}}$), the X-ray bolometric correction ($\kappa_{\rm bol}^{\rm X}=L_{\mathrm{bol}}/L_{\mathrm{2-10\,keV}}$) and the optical-to-X-ray spectral slope ($\alpha_{\rm OX}$), for the whole sample. These latter two parameters are commonly used in AGN studies to characterise the relative contributions of UV and X-ray emission produced by the accretion disc and the hot corona, respectively. The $\alpha_{\rm OX}$ parameter differs from $\kappa_{\rm bol}^{\rm X}$ in that it concentrates on the monochromatic luminosities at rest--frame 2500\,\AA\ and 2\,keV, and is defined as $\alpha_{\rm OX}=\log(L_{2\,\rm keV}/L_{2500\,\AA})/\log(\nu_{2\,\rm keV}/\nu_{2500\,\AA})$.  To compute $\kappa_{\rm bol}^{\rm X}$ and $\alpha_{\rm OX}$ we adopt the values derived from the power-law fits.\\
\indent For J183-12, which is not detected, and for J0313-1806, which is marginally detected but with a number of net counts $\sim40$\% below the required threshold (see Sect.~\ref{subsect:source_detection}), we computed EPIC-PN upper limits on the fluxes, luminosities, $\alpha_{\rm OX}$, and $\kappa_{\rm bol}^{\rm X}$. These limits are listed in Tab.~\ref{table:sources}, together with the spectral measurements for the X-ray detected objects. In particular, we estimated the fluxes with \texttt{xspec}, using the spectral response files extracted at the source positions and adopting a power-law model with photon indices $\Gamma=2$ and $\Gamma=2.4$ (the latter corresponding to the mean $\Gamma$ of the sample, as reported in \citetalias{Zappacosta2023}), both attenuated by the Galactic column density. We then computed the intrinsic (i.e. unabsorbed) 2–10~keV and 2~keV luminosities with \texttt{xspec}, assuming the same power-law spectral model.\\
\indent Since J0313-1806 is the most distant hyperluminous QSO known so far ($z=7.64$, see Tab.~\ref{table:sources}), the only one with available mass measurement, it is worth performing an exploratory spectral analysis based on its marginal detection. We found a lower limit to the photon index $\Gamma >1.77$ owing to the limited photon statistics. We note that the presence of two neighbouring sources within $\sim20$ arcsec from J0313-1806 may significantly impact on the determination of the $\Gamma$ value and on the X-ray properties reported in Tab.~\ref{table:sources}.\\
\indent The analysis of the remaining HYPERION X-HP observations of three QSOs (J083, J0038 and J036) allows us to finalise the X-ray study of the full HYPERION sample (see Tab.~\ref{table:sources}). For the X-ray spectral datasets of the new X-LP QSOs, we find that all detected sources display steep spectral slopes, in good agreement with previous HYPERION results \citepalias{Zappacosta2023,Tortosa2024}.  The mean $\Gamma$ of the complete HYPERION sample is $\langle\Gamma\rangle_{\rm HYP}=2.46$, the standard deviation of the $\Gamma$ distribution is $\sigma_{\rm HYP}=0.39$; the value obtained for the full sample analysed in this work is $\langle\Gamma\rangle=2.40$ and its standard deviation is $\sigma=0.38$. When considering only the non-HYPERION QSOs, including the new X-LP targets, the average $\langle\Gamma\rangle_{\rm non-HYP}$ is $2.27$ and the standard deviation is $\sigma_{\rm non-HYP}=0.34$. This steep X-ray continuum curvature could equivalently be reproduced by a low high-energy cutoff ($E_{\rm c}$), lying well within the observed energy band and corresponding to $<50$\,keV in the rest-frame \citepalias{Zappacosta2023,Tortosa2024}. In the presence of limited quality spectra such as those we are reporting for EoR QSOs, it is not possible to distinguish between intrinsically steep power-laws and canonical power-laws ($\Gamma=1.9$) with a relatively low value of the cutoff energy ($E_{\rm c}\sim20$\,keV). We measured $E_{\rm c}$ also in the new dataset by modelling the spectra with a cutoff power-law model (\texttt{const}*\texttt{tbabs}*\texttt{zcutoffpl} in \texttt{xspec}), fixing the photon index to $\Gamma=1.9$. New values together with the values of previously analysed datasets are reported in Tab.~\ref{table:sources}. The observed frame 2--10~keV fluxes and the rest-frame 2--10~keV luminosities inferred from the power-law model are consistent, within the statistical uncertainties, with those obtained from the cutoff power-law model with the photon index fixed at $\Gamma=1.9$. Since no significant differences are found, we adopt the values derived from the power-law fits throughout this work. We note that $E_{\rm c}$ values should be regarded as an alternative parameterisation of the observed spectral curvature rather than an independent measurement of the coronal cutoff energy, which should be kept in mind when comparing them with lower-redshift samples where $\Gamma$ and $E_{\rm c}$ were constrained simultaneously. We further note that a few of the lowest fitted values (e.g. $E_{\rm c} \approx 3.9$~keV for DES~J0252-0503) lie close to the lower edge of the adopted rest-frame fitting range ($\sim2$~keV), reinforcing the need for caution in interpreting them as physical coronal cutoff energies.
\subsection{Correlation analysis}
\label{sect:correl_analysis}
\begin{figure*}
\centering
\includegraphics[width=0.9\columnwidth]{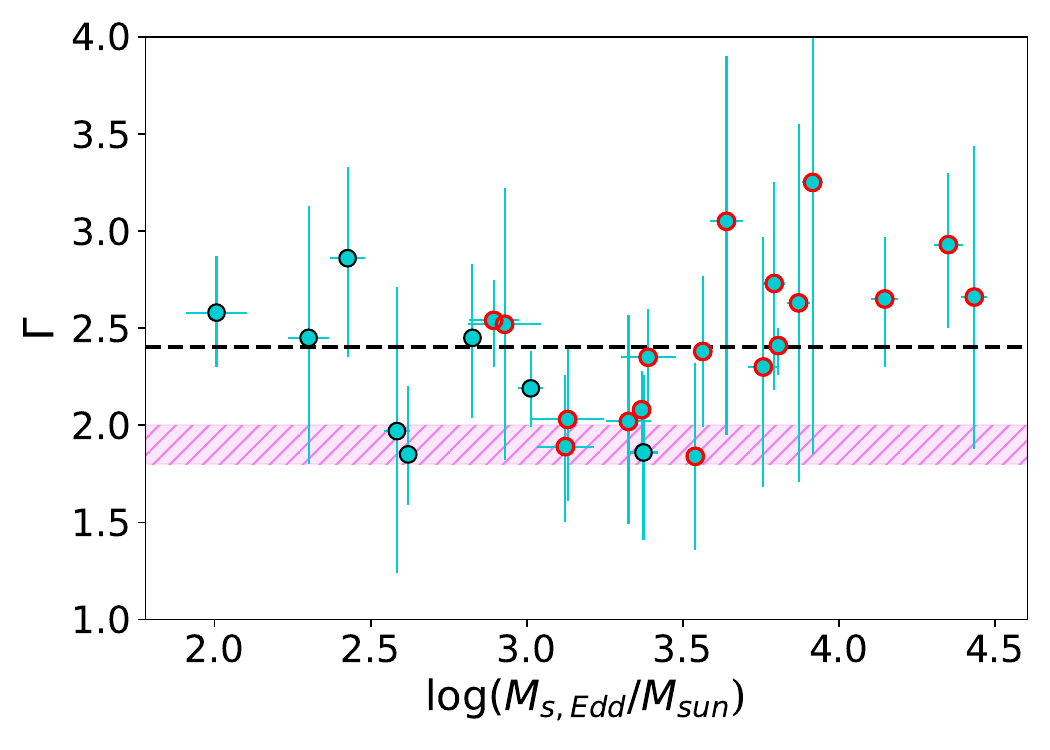}
\includegraphics[width=0.9\columnwidth]{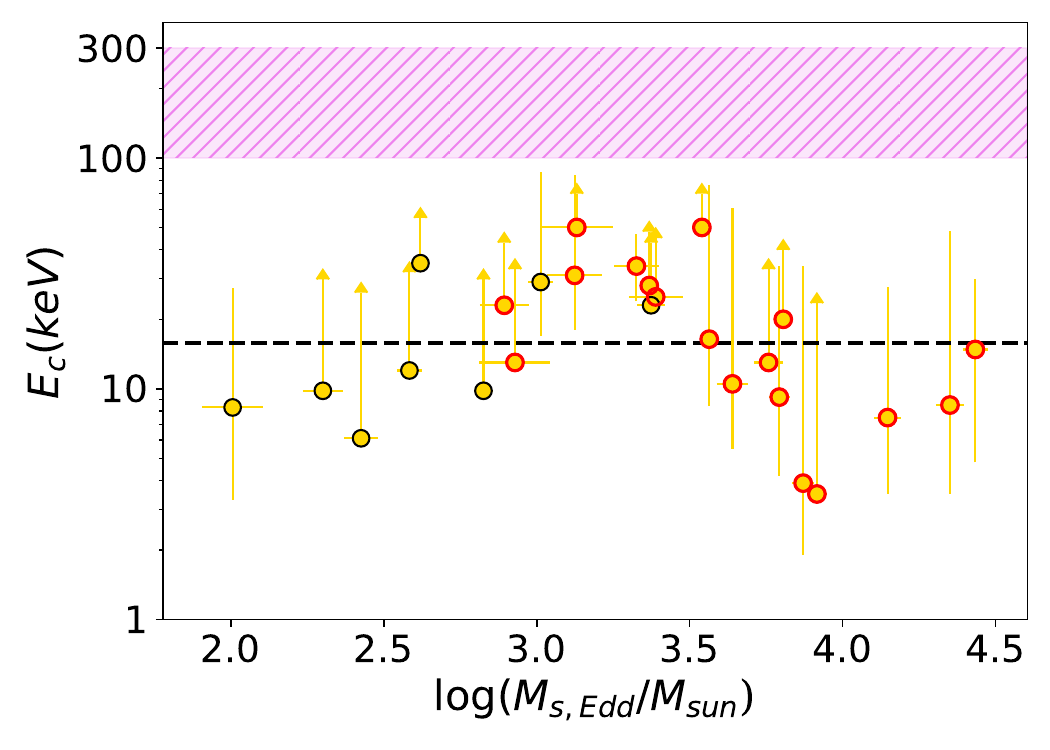}
\caption{\textit{Left panel}: plot of $\Gamma$ vs. \mseed; \textit{Right panel}: plot of $E_{\rm c}$ vs \mseed. The black horizontal dashed line marks the mean value of $\Gamma$ and $E_{\rm c}$ for the sample of detected QSOs. Points with red edges are the ones belonging to the HYPERION sample. Violet shaded region marks the typical range of $\Gamma$ ($\sim1.8-2$; e.g. \citealt{Piconcelli2005,Ricci2017}) and $E_{\rm c}$ ($\sim100-300$\,keV; e.g. \citealt{Fabian2015,Ricci_2018,Lanzuisi19}) observed in low-$z$ unobscured AGN. We note that the \mseed\ values plotted here are, for simplicity, the mean values derived using all the available values reported in Tab.~\ref{table:masses}.}
\label{fig:mseed_plots}
\end{figure*}
The presence of a possible linear relation between pairs of quantities was investigated using the \texttt{linmix} code, a hierarchical Bayesian model for fitting $y=Ax+B$ accounting for errors in $x$ and $y$ \citep{Kelly_2007} assuming symmetric Gaussian uncertainties. To account for asymmetric uncertainties and for the multiple available estimates of some quantities for the same source, we embedded the fit in a Monte Carlo procedure of $10^4$ realisations of the fit to propagate the observational uncertainties. At each iteration, every point was resampled according to its nature: (i) for a single measurement with a (possibly asymmetric) uncertainty, a value was drawn from a Gaussian, or an asymmetric split-normal distribution when the lower and upper errors differ (see App.~\ref{app:gamma_distribution} and Eq.~\ref{eq:distribution}); (ii) for a quantity subject to an upper or lower limit, we applied the censored fitting method (CF; \citealt{Guainazzi2006,Bianchi2009}), substituting each upper (lower) limit $U$ ($L$) with a value drawn from a uniform distribution in $[\#_{\rm min},U]$ ($[L,\#_{\rm max}]$), with $\#_{\rm min}\ll U$ ($\#_{\rm max}\gg L$). Among the detected sources included in the fits, this applies only to the 15 lower limits on the high-energy cutoff, for which we adopted $E_{\rm c,max}=500$~keV, well above both the largest lower limit in our sample ($E_{\rm c}>50$~keV) and the cutoff energies typically measured in local AGN \citep[e.g.][]{Ricci_2018}. Since 15 of the 26 $E_{\rm c}$ values are lower limits, we verified that this result does not depend on the adopted censored-fitting assumptions. Repeating the full Monte Carlo procedure with $E_{\rm c,max}=200$ and $1000$~keV, and drawing the substituted values from a log-uniform instead of a uniform distribution, yields $\rho_P$ between -0.25 and -0.18 ($<1\sigma$) in all cases.\\
\indent The investigation of the relations involving the \mseed\ parameter critically depends on the multiple values of $M_{\rm BH}$ (see App.~\ref{app:masses}), often derived using different single epoch virial estimators (based on Mg~\textsc{ii}, H$\alpha$ and H$\beta$ emission lines). For each source, whenever multiple $M_{\rm BH}$ measurements obtained with the same virial estimator were available, we computed a representative black hole mass by adopting their mean value among the ones we re-derived using consistent scaling relations and cosmological assumptions, $\langle M_{\rm BH} \rangle$. The uncertainties associated with the mean values are given by the distance to the minimum/maximum. If only one measurement was available for a given estimator, that value and its uncertainty were adopted directly. For each realisation we computed the \texttt{linmix} fit, its $\chi^2$, the Pearson coefficient $\rho_P$, and the null-hypothesis probability $P_{\rm null}$; to prevent atypical draws (e.g. several sources scattered toward the same tail) from skewing the result, we retained only the realisations with $\chi^2$ within $\pm1\sigma$ of the peak of its distribution over all $10^4$ iterations, and report the mean slope, intercept, $\rho_P$ and $P_{\rm null}$ over these realisations. $P_{\rm null}$ is computed analytically from $\rho_P$ via the Student's $t$-distribution for $N$ points, and converted into an equivalent significance as $\sigma=\Phi^{-1}(1-P_{\rm null}/2)$. Undetected sources (i.e. J~183-12 and J~0313-1806, see Sect.~\ref{subsect:source_detection}) are excluded a priori from all fits.\\
\begin{figure}
    \centering
    \includegraphics[width=0.9\columnwidth]{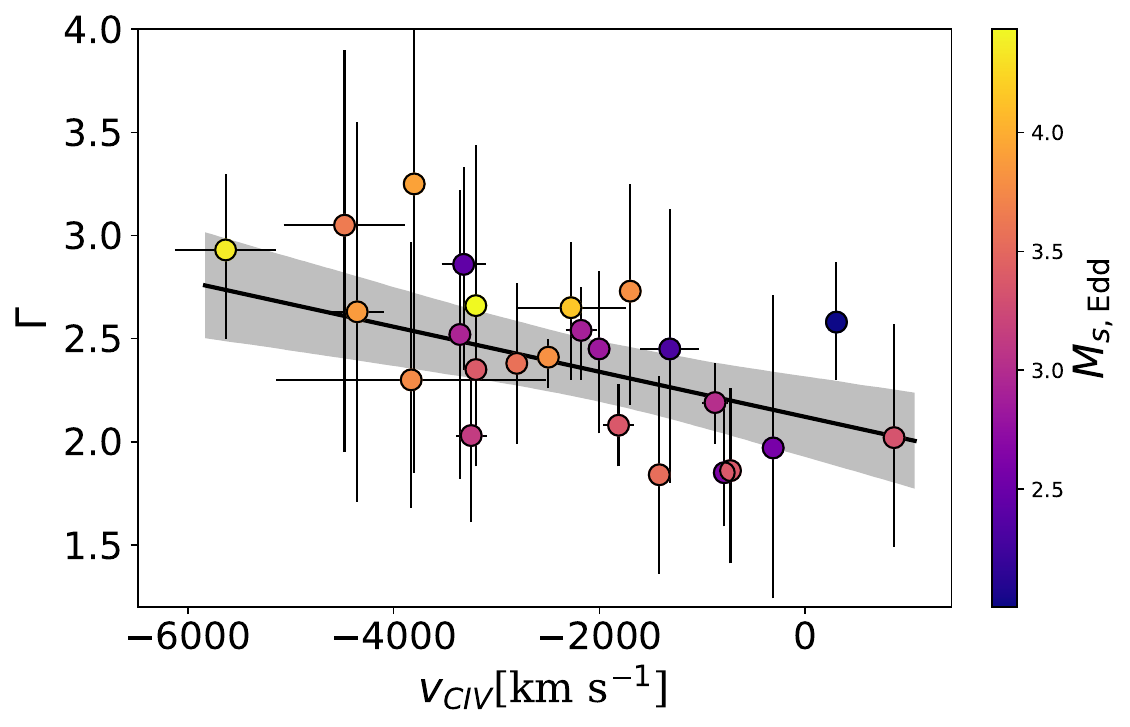}
    \caption{$\Gamma$ as a function of $v_{\rm C~\textsc{iv}}$. Points are colour coded for \mseed. The horizontal dashed line represents the mean value of $\Gamma$. The solid black line indicates the best fitting relation, with the shaded region showing its $1\sigma$ uncertainty }
    \label{fig:vciv_gamma}
\end{figure}
\indent We investigated how the coronal X-ray properties are linked to \mseed\ and therefore relate to SMBH growth. We first examined the relation between \mseed\ and the X-ray coronal spectral parameters, namely $\Gamma$ and $E_{\rm c}$ (see Fig.~\ref{fig:mseed_plots}). The left panel of Fig.~\ref{fig:mseed_plots} displays $\Gamma$ as a function of \mseed. We find a Pearson correlation coefficient of $\rho_{P} = 0.15$ and a null-hypothesis probability of $P_{\mathrm{null}} = 0.44$, corresponding to a statistical significance of $0.8\sigma$. No statistically significant correlation between $\Gamma$ and \mseed\ is found in our sample, in contrast with the marginally positive correlation previously reported by \citetalias{Tortosa2024}. The detected newly observed QSOs from the X-LP, covering the lowest range of \mseed, continue to exhibit steep slopes (see Tab.~\ref{table:sources}). We additionally tested the robustness of this result by progressively removing the most extreme sources in \mseed, repeating the full Monte Carlo procedure described above on the reduced sample. Removing one source per side ($N=24$) yields $[\rho_{P};P_{\mathrm{null}}]=[0.18;0.40]$ ($0.9\sigma$); removing two sources per side ($N=22$) yields $[0.17;0.41]$ ($0.8\sigma$); removing three sources per side ($N=20$) yields $[0.24;0.33]$ ($1.0\sigma$). The correlation remains weak at all steps, indicating that this result is not driven by a small number of outliers. The posterior slope, $A=0.05\pm0.11$ ($\Gamma\,{\rm dex}^{-1}$), excludes at the $\sim95\%$ level relations steeper than $|A|\gtrsim0.27\,\Gamma\,{\rm dex}^{-1}$, corresponding to a change in $\Gamma$ larger than $\sim0.65$ over the $\sim2.4$~dex in $\log(M_{\rm s,Edd})$ spanned by our sample; weaker relations remain statistically indistinguishable from the null result given the precision of the current data. The right panel of Fig.~\ref{fig:mseed_plots} shows $E_{\mathrm{c}}$ as a function of \mseed. Here, the Pearson coefficient is $\rho_{P} = -0.24$ with $P_{\mathrm{null}} = 0.30$, corresponding to a significance of $1.0\sigma$. Progressively removing the most extreme sources in \mseed\ yields $[\rho_{P};P_{\mathrm{null}}]=[-0.32;0.21]$ ($1.3\sigma$, $N=24$), $[-0.25;0.33]$ ($1.0\sigma$, $N=22$), and $[-0.17;0.46]$ ($0.8\sigma$, $N=20$). No statistically significant correlation is therefore present between $E_{\rm c}$ and \mseed\ in our sample, and this result is stable against the exclusion of the most extreme sources. In summary, after enlarging the sample to include the X-LP targets, neither the $\Gamma$ nor $E_{\rm c}$ exhibits a statistically robust dependence on \mseed. We further investigated the dependence of the broadband and X-ray luminosity properties on \mseed\ (see App.~\ref{app:other_relations}). The results of the linear regression analysis are summarised in Tab.~\ref{table:all_fitting_param}. None of these quantities show a statistically significant relation with \mseed. We also verified that $\Gamma$ shows no significant dependence on $z$ or $M_{\rm BH}$ individually ($\rho_P=0.21$, $1.0\sigma$, and $\rho_P=-0.11$, $0.6\sigma$, respectively), nor when their effects are disentangled through partial correlation and multiple regression analysis (both $\lesssim1\sigma$), consistent with the spectral steepness found in our sample not being primarily driven by either the redshift range or the black hole mass distribution probed here.\\
\indent We also explored the connection between the coronal X-ray properties and the broad-line wind velocity traced by the C~\textsc{iv} velocity shift relative to the systemic redshift, $v_{\rm C~\textsc{iv}}$. Fig.~\ref{fig:vciv_gamma} shows $\Gamma$ vs. $v_{\rm C~\textsc{iv}}$, while Fig.~\ref{fig:vciv_plots} shows $L_{\rm 2-10\,keV}$ and $\alpha_{\rm OX}$ vs. $v_{\rm C~\textsc{iv}}$. Following \citetalias{Tortosa2024}, we accounted for multiple $v_{\rm C~\textsc{iv}}$ measurements reported in the literature. Specifically, we repeated the fitting procedure 10,000 times, and at each iteration we randomly selected one of the available $v_{\rm C~\textsc{iv}}$ values for QSOs with multiple measurements, adopting the same methodology described above for \mseed. The $\Gamma - v_{\rm C~\textsc{iv}}$ correlation reported by \citetalias{Tortosa2024} remains present ($>3\sigma$) in the expanded sample (see Fig.~\ref{fig:vciv_gamma}). In contrast, neither the X-ray luminosity nor the $\alpha_{\rm OX}$ exhibits a statistically significant dependence on $v_{\rm C~\textsc{iv}}$ at variance with what is found at Cosmic Noon ($z\sim2-4$) for analogous QSOs with similar luminosities and Eddington ratio which clearly show a trend for decreasing X-ray luminosities for increasing $v_{\rm C~\textsc{iv}}$ \citep{Zappacosta2020,2025A&A...702A.114D}.
\section{Discussion}
\label{sect:Discussion}
\subsection{X-ray spectral steepness}
\label{subsec:x-rayspectralsteepness}
Our current analysis, which relies on the most extensive and sensitive X-ray spectroscopic study of QSOs up to $z=7.6$, does not recover a statistically significant correlation between $\Gamma$ and \mseed\, at variance with the tentative evidence for such a link reported in \citetalias{Tortosa2024}. The QSOs in our sample display, on average, steep X-ray continua, and this finding suggests that spectral steepness is likely an intrinsic property of luminous QSOs at the EoR, rather than a consequence of selection effects tied to mass growth, which could have biased the HYPERION sample. This result is further supported by a dedicated statistical analysis, in which we reconstruct the uncertainty-propagated distribution of the measured $\Gamma$ values via Monte Carlo simulations, confirming that our sample is systematically characterised by steep X-ray spectra (see App.~\ref{app:gamma_distribution} for the details). As shown in Fig.~\ref{fig:gamma_mc}, the reconstructed distribution remains centred at $\Gamma\sim2.4$ and is significantly shifted toward steeper values than that observed in a comparison sample (Fig.~\ref{fig:gamma_mc}) of WISE/SDSS-selected Hyper-luminous QSOs (WISSH) at Cosmic Noon ($z\sim2-4$) from \citealt{2025A&A...702A.114D} with similar $M_{\rm BH}$ and $L_{\rm bol}$, whose distribution peaks around the canonical $\Gamma\sim1.9$.\\
\indent A two-sample Kolmogorov--Smirnov test confirms that the two distributions are statistically different, both when applied to the best-fit values ($D=0.60$, $1-P=0.999$) and over $10\,000$ Monte Carlo realisations accounting for the measurement uncertainties (median $1-P=0.9906$, with $1-P>0.95$ in $\sim98\%$ of the realisations; App.~\ref{app:gamma_distribution}). This difference is not driven by the different rest-frame bands probed in the two samples: restricting the spectral analysis of WISSH QSOs with comparable spectral quality to rest-frame energies above 2~keV, i.e. the same lower bound probed for our $z\gtrsim6$ sources, still yields canonical photon indices \citepalias{Zappacosta2023}.\\
\indent The fact that WISSH QSOs generally exhibit canonical X-ray slopes \citep[e.g.][]{2025A&A...702A.114D} could indicate that this spectral steepness is not a simple consequence of luminosity or black-hole mass, but instead points to different physical conditions in the accretion flow during the earliest phases of SMBH growth. Although this conclusion is currently limited to the luminous QSOs population, the lack of a significant dependence of $\Gamma$ on AGN luminosity observed at lower redshift \citep[e.g.][]{Piconcelli2005,Ricci2017} suggests that a similar behaviour for EoR AGN with lower luminosities is plausible, although it remains to be tested observationally. As a consistency check, we verified that the steep $\Gamma$ values are not driven by a single instrument: the $\Gamma$ distributions of the \xmm-observed ($N=23$) and \textit{Chandra}-observed ($N=3$) subsamples are statistically indistinguishable (two-sample KS test, $D=0.54$, $p=0.32$), although the small size of the \textit{Chandra} subsample limits the power of this test. Moreover, a stacked \textit{Chandra} spectrum of 11 QSOs at $z>6.5$, partly overlapping with our sample, yields $\Gamma=2.41\pm0.27$ \citep{Jin2026}, in agreement with our mean value.
\begin{figure}
    \centering
    \includegraphics[width=0.9\columnwidth]{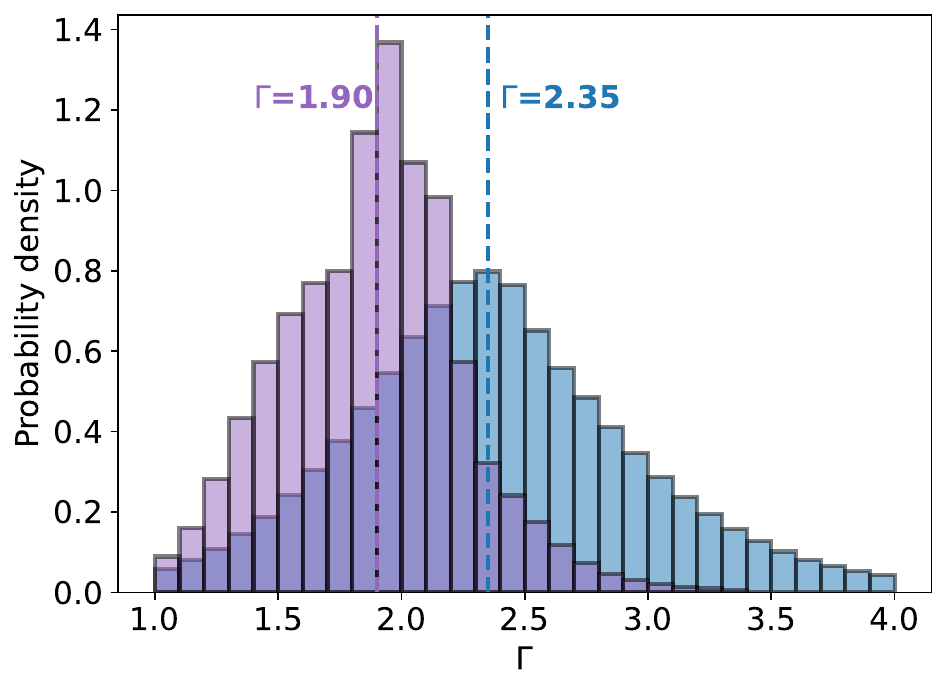}
    \caption{Probability density distributions of $\Gamma$ for the $z \gtrsim 6$ QSOs (blue) from this work and for the $z\sim2-4$ QSOs samples (purple) from \citealt{2025A&A...702A.114D}. Dashed lines represent the median value of the distributions (see Tab.~\ref{tab:gamma_distribution}).}
    \label{fig:gamma_mc}
\end{figure}
We further investigated whether the few QSOs with canonical photon indices could instead be intrinsically steep-spectrum sources affected by moderate intrinsic absorption. The results of these tests are presented in App.~\ref{app:test_steep_spectra}. Absorbed models are not statistically preferred over a simple unabsorbed power-law, the current data quality does not allow us to establish whether these apparently flat-spectrum QSOs are intrinsically steep but moderately obscured systems.\\
\indent Owing to the well-known degeneracy between $\Gamma$ and $E_{\rm c}$ in broadband X-ray spectral modelling \citep[e.g.][]{Dadina08,2018A&A...614A..37T}, both these two properties may contribute to the spectral steepness observed in our sample of $z \gtrsim 6$ QSOs. The uncertainty-propagated distribution of the high-energy cutoff cannot be reconstructed using Monte Carlo simulations in the same way as for $\Gamma$. Unlike the photon index, which is directly measured for all the targets, a significant fraction of the QSOs in our sample provides only lower limits on $E_{\rm c}$, preventing a meaningful reconstruction of its uncertainty-propagated distribution. Moreover, demonstrating that uncertainty-propagated distribution of $\Gamma$ is systematically shifted toward high values is sufficient to establish that the X-ray continua of luminous $z \gtrsim 6$ QSOs are intrinsically steeper than those of lower-$z$ AGN, regardless of whether the spectral curvature is parametrised through a steep photon index or a low high-energy cutoff.
\begin{figure}
    \centering
    \includegraphics[width=0.9\columnwidth]{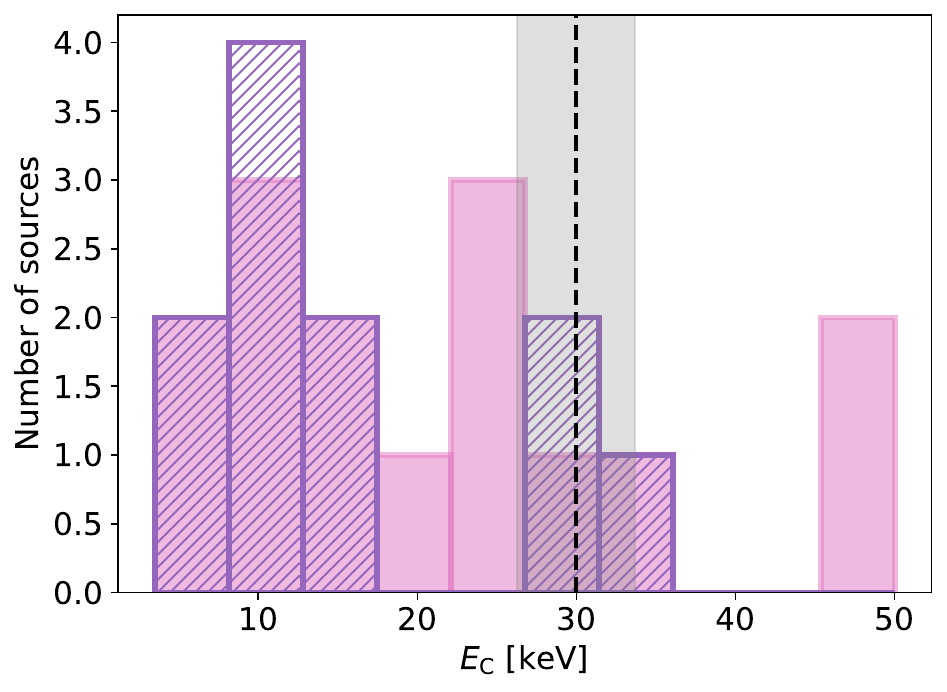}
    \caption{High-energy cutoff distribution. Detections are shown in hatched purple, while lower limits are shown in pink. The dashed vertical line marks the restricted mean cutoff energy derived from the Kaplan--Meier estimator $\langle E_{\rm c}\rangle_{\rm KM}=30.0\pm3.7$\,keV.}
    \label{fig:Ec_distrib}
\end{figure}
We show the $E_{\rm c}$ distributions of the sources in our sample in Fig.~\ref{fig:Ec_distrib}. Due to the high fraction of lower limits on $E_{\rm c}$ using a simple arithmetic mean is not appropriate. To include the lower limits, we used survival analysis (SA; e.g., \citealt{1985ApJ...293..192F,2017MNRAS.466.3161S}), computing with \texttt{scikit-survival} \citep{sksurv} the non-parametric Kaplan-Meier product-limit (KMPL) estimator of the survival function (1-CDF), which requires no parametric assumption and is consistent under non-informative censoring. Using the KMPL, we calculated the restricted mean $E_{\rm c}$ (i.e. the mean truncated at the largest value, detected or censored, present in the sample, $\tau=50$~keV in our case) and estimated its uncertainty, since the survival function does not reach zero within the observed range and the unrestricted mean is therefore not defined. The KMPL estimator requires no parametric distributional assumption and, under non-informative censoring, provides a consistent estimate of the survival function. The uncertainty on the restricted mean was propagated from the Kaplan--Meier variance up to $\tau$. Accounting for the presence of 15 lower limits using the Kaplan–Meier estimator, we derived a restricted mean cutoff energy of $\langle E_{\rm c}\rangle_{\rm KM}=30.0\pm3.7$ keV. Our survival-analysis estimate is substantially lower than the mean value reported for similar QSOs at lower redshift, e.g. the SUBWAYS QSOs ($0.1 < z < 0.4$) have a mean $E_{\rm c}\sim87$ keV \citep{Peluso2026}, which is already at the low end of the distribution observed in local Seyfert galaxies ($E_{\rm c}\sim100-200$\,keV). Luminous QSOs at the EoR thus may host even colder and/or more optically thick coronae than those observed in luminous QSOs at lower-$z$.
\begin{figure}
\centering
\includegraphics[width=1.\columnwidth]{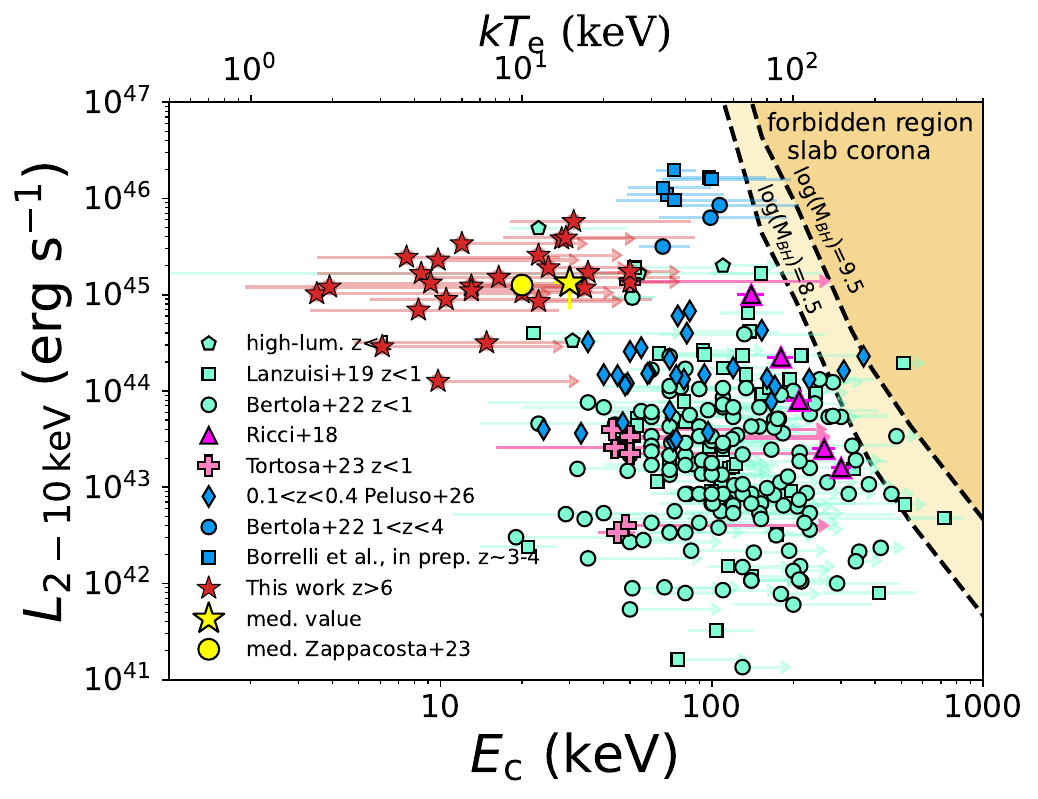}
\caption{$L_{2-10\,\rm keV}$ versus $E_{\rm c}$. The upper x-axis shows the corresponding electron temperature, $kT_e$ assuming a slab corona. Sources analysed in this work are the red stars. We report also a compilation of values from the literature: binned average of the values of the BASS survey from \citealt{Ricci_2018} (magenta triangles), $z<1$ values collected by \citealt{Lanzuisi19} (turquoise squares), $z<1$ values by \citealt{Bertola22} (turquoise circles), $z<1$ high-luminous AGN \citep[][turquoise pentagons]{Kammoun2017,Reeves2021,Marinucci2022,Kammoun2023,Lanzuisi2024}, $0.1<z<0.4$ values from \citealp{Peluso2026} (azure diamonds), $1<z<4$ values collected by \citealt{Bertola22} (azure circles), $1<z<3$ values from Borrelli et al. in prep. (azure squares), $z<1$ highly accreting AGN from \citealt{Tortosa2023} (pink crosses). Yellow star is the median values of the sources considered in this work and the yellow circle is the median values from \citetalias{Zappacosta2023}. Yellow regions are the forbidden regions (for a slab corona model) due to runaway electron-positron pair production \citep[see][]{1984MNRAS.209..175S} for $\log(M_{\rm BH}/M_{\odot})=8.5$ and $\log(M_{\rm BH}/M_{\odot})=9.5$.}
\label{fig:Ec_Lx}
\end{figure}
Fig.~\ref{fig:Ec_Lx} presents the cutoff energy as a function of the 2--10~keV luminosity. The QSOs analysed in this work are compared with many different samples from the literature, spanning a broad range of redshift, luminosity, and accretion regimes. Compared to lower-$z$ AGN samples, our sources tend to populate the high-$L_{2-10\,\rm keV}$ and low-$E_{\rm c}$ region of the diagram, implying systematically colder coronae as $E_{\rm c}$ scales with the coronal electron temperature, $kT_{\rm e} \sim E_{\rm c}/2-3$ depending on the adopted corona geometry \citep{Petrucci2000,Petrucci2001}. They differ significantly also from the limited number of measurements available for QSOs with comparable $L_{2-10\,\rm keV}$. However, they are compatible with a few measurements in highly accreting AGN. \\
\indent The pervasive X-ray spectral steepness observed in luminous EoR QSOs, whether parametrised as a steep $\Gamma$, a low $E_{\rm c}$, or a combination of both, points to systematically colder coronae than those typically observed in lower-$z$ AGN. The physical origin of this coronal state, however, remains unclear. One possibility is that these systems intrinsically host lower electron temperatures than those of lower-$z$ QSOs. Since the physical mechanism responsible for coronal heating is still uncertain, involving processes such as magnetic reconnection and magnetic turbulence generated by the magneto-rotational instability, differences in the magnetic or plasma properties of the accretion flow could naturally lead to lower electron temperatures and hence steeper X-ray continua \citep[see][for a recent review]{2025FrASS..1130392L}. Alternatively, the coronae may not be intrinsically different, but rather experience more efficient radiative cooling due to a stronger field of UV and soft X-ray seed photons produced by the accretion flow. Such a scenario is expected in systems accreting at very high Eddington ratios \citep[e.g.][]{2017MNRAS.468.3489K,Tortosa22,Tortosa2023}, where enhanced Compton cooling naturally lowers the coronal temperature, producing steeper X-ray spectra with low high-energy cutoffs. In this respect, luminous QSOs at the EoR may share similarities with local narrow-line Seyfert 1 galaxies \citep{Berton2026,Tortosa2026}, which also exhibit steep X-ray spectra and, in several cases, relatively low $E_{\rm c}$.\\
\indent Interestingly, although the EoR QSOs analysed here are several orders of magnitude more luminous and massive than the local super-Eddington AGN studied by \citet{Tortosa2023}, both populations occupy the low-$E_{\rm c}$ regime ($E_{\rm c}\lesssim50$\,keV; see Fig.~\ref{fig:Ec_Lx}). This suggests that similarly low coronal temperatures can be reached over a remarkably broad range of black-hole masses and luminosities, possibly through the same physical mechanism, namely enhanced Compton cooling driven by the intense radiation field associated with extreme accretion. Thus, the overall observational picture, including the $\Gamma - v_{\rm C~\textsc{iv}}$ correlation, which remains present in the expanded sample, is qualitatively consistent with the scenario proposed by \citetalias{Tortosa2024}, in which the X-ray corona is efficiently cooled by the intense radiation field associated with high accretion regimes. Future high-throughput missions extending to lower observed energies will be crucial to test whether luminous EoR QSOs also exhibit enhanced soft X-ray emission capable of efficiently cooling the corona.
\subsection{X-ray-to-UV properties and bolometric corrections}
\label{subsec:X-rayandUV}
\begin{figure*}
    \centering
    \includegraphics[width=0.9\columnwidth]{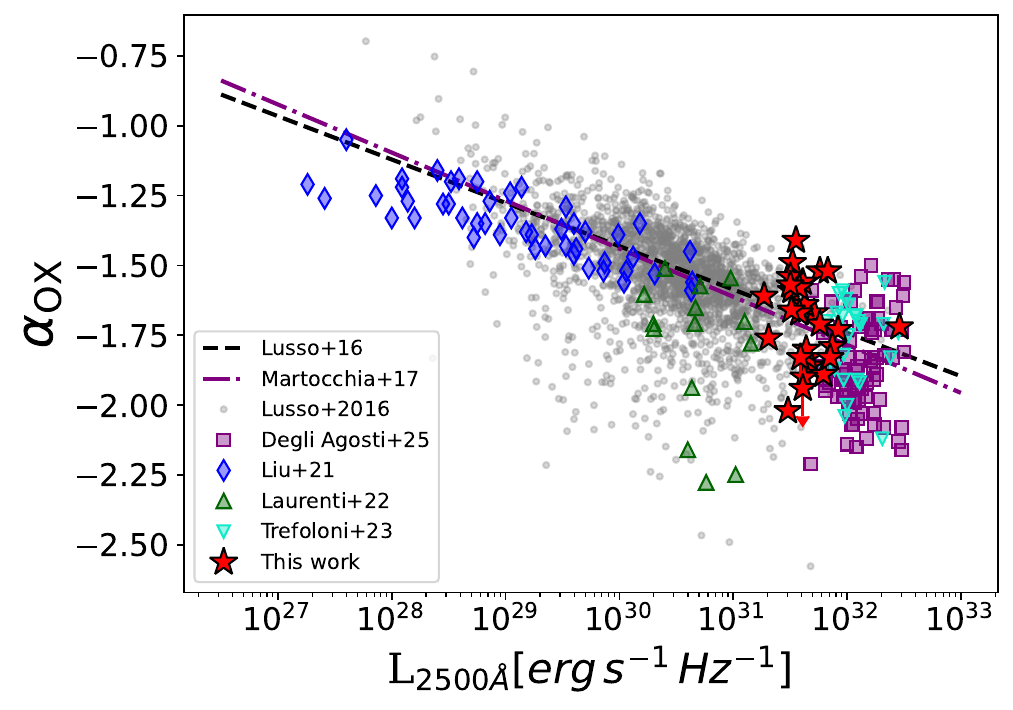}
    \includegraphics[width=0.9\columnwidth]{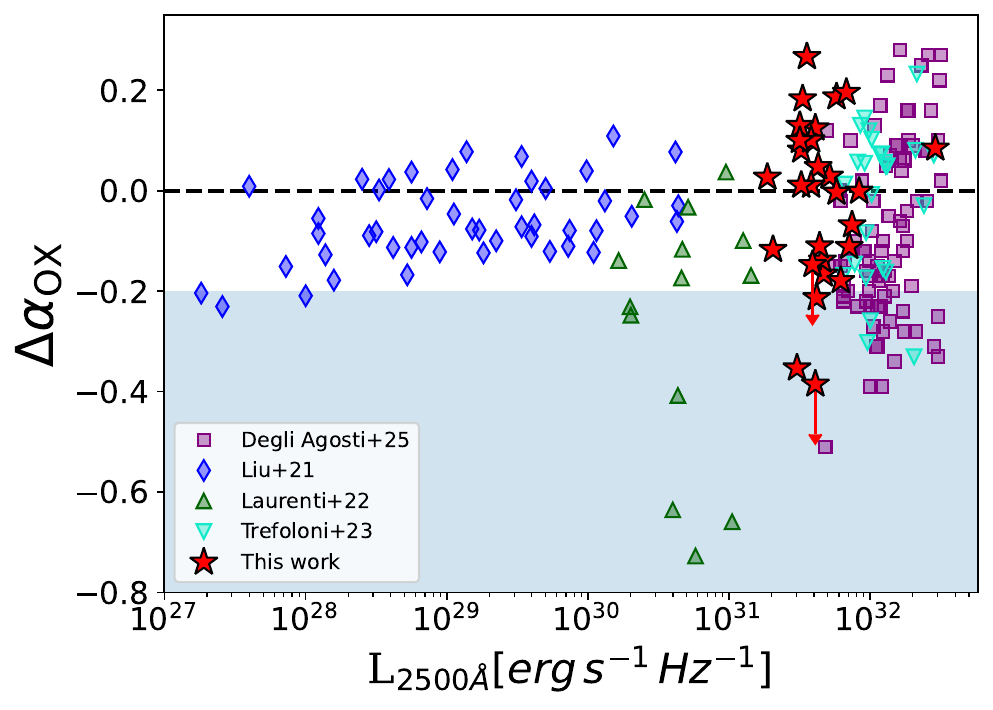}
    \caption{Left panel:$\alpha_{\rm OX}$ as a function of $L_{\rm 2500}$. Black dashed line refers to the best-fit relation from \citealt{Lusso2016}, purple dotted-dashed line refers to the best-fit relation from \citealt{Martocchia2017}; Right Panel: $\Delta(\alpha_{\rm OX}) = \alpha_{\rm OX} - \alpha_{\rm OX, RL16}$ as a function of $L_{\rm 2500}$; blue shaded region includes the X-ray weak QSOs \citep{Luo_2015}. In all the panels the sources analysed in this work are the red stars. We plot for comparison the following samples: X-ray selected type 1 AGN in XMM-COSMOS $0.04<z<4.5$ from \citealt{Lusso2016} (gray points), WISSH QSOs at $z\sim 2-4$ from \citealt{2025A&A...702A.114D} (purple squares), local AGN/NLS1s from \citealt{2021ApJ...910..103L} (blue diamonds), nearby $0.40<z<0.75$ highly accreting AGN from \citealt{2022A&A...657A..57L} (dark green triangles), SDSS DR7 QSOs at $z\simeq 3.0-3.3$ from \citealt{2023A&A...677A.111T} (turquoise inverted triangles).}
    \label{fig:aox_LUV}
\end{figure*}
We first investigated the $\alpha_{\rm OX}$ parameter. The left panel of Fig.~\ref{fig:aox_LUV} displays several AGN samples together with the best-fit relations reported by \citealt{Lusso2016} and \citealt{Martocchia2017}. In the $\alpha_{\rm OX}$–$L_{2500}$ plane, the $z \gtrsim 6$ QSOs analysed in this work are distributed both above and below the relations derived at lower-$z$. This behaviour can be further quantified through the deviation between the observed $\alpha_{\rm OX}$ and the value expected from the UV luminosity according to the \citealt{Lusso2016} relation (LR16), defined as $\Delta\alpha_{\rm OX} = \alpha_{\rm OX}(\rm observed) - \alpha_{\rm ox,LR16}(L_{2500\AA})$. The right panel of Fig.~\ref{fig:aox_LUV} shows $\Delta\alpha_{\rm OX}$ as a function of the 2500~\AA\ luminosity. \\
\indent adopting $\Delta\alpha_{\rm OX}<-0.2$ as the threshold for classifying X-ray-weak sources \citep{Luo_2015}, we estimated the incidence of X-ray weakness in our sample. For the archival non-HYPERION QSOs, the $>30$ net-count requirement (Sect.~\ref{sect:sample_and_data_red}) depends jointly on the intrinsic X-ray flux and on the exposure depth, and may therefore disfavour intrinsically X-ray-weak sources in a way that is difficult to quantify. We thus restricted this estimate to the 23 HYPERION and X-LP QSOs, which were selected on the basis of their rest-frame UV and SMBH properties, independently of their X-ray properties, and targeted with dedicated \xmm\ programmes. None of these sources is excluded: for the two QSOs below the count threshold (J0313-1806 and PSO~J183-12; Sect.~\ref{subsect:source_detection}) we used the EPIC-PN upper limits. Two detected QSOs, PSO~J025-11 ($\Delta\alpha_{\rm OX}=-0.35$) and PSO~J231.6-20.8 ($-0.21$), satisfy the threshold, as does J0313-1806, whose upper limit ($\Delta\alpha_{\rm OX}\lesssim-0.38$ and $\lesssim-0.35$ for $\Gamma=2$ and $2.4$, respectively) lies well below it. This yields an X-ray-weak fraction of 3/23, i.e. $13^{+13}_{-7}\%$ (1$\sigma$; \citealt{Gehrels1986}), with a 90\%-confidence upper bound of $\sim29\%$. The upper limit for PSO~J183-12 ($\Delta\alpha_{\rm OX}\lesssim-0.15$), whose observation was strongly affected by background flaring (Tab.~\ref{table:obsid}), does not allow a classification; should this source also be X-ray weak, the fraction would rise to 4/23 ($17^{+14}_{-8}\%$). We note that PSO~J231.6-20.8 lies just below the threshold, while J1007+2115 and DELS~J0038-1527 ($\Delta\alpha_{\rm OX}=-0.17$ and $-0.18$) lie just above it, so that the classification of these borderline sources is sensitive to the uncertainties on $\alpha_{\rm OX}$. For reference, none of the archival non-HYPERION QSOs is X-ray weak, so that the fraction over the full sample is 3/28 ($\sim11\%$); however, this value is subject to the count-based selection discussed above. The fraction derived for the HYPERION and X-LP QSOs appears lower than, but not statistically inconsistent with, those reported for other samples of highly accreting QSOs, for which X-ray weakness fractions of $\sim25-30\%$ have been found \citep[e.g.][]{Nardini2019,2023A&A...677A.111T,2020MNRAS.499.6053L,2022A&A...657A..57L,2025A&A...702A.114D}, a range that overlaps with our 90\%-confidence upper bound.\\
\indent We also examined the X-ray to bolometric luminosity correction. We compared the $\kappa^{x}_{\rm bol}$ values of our sample with those derived for several samples available in the literature. Fig.~\ref{fig:kbol_Lbol} shows the bolometric correction $\kappa^{x}_{\rm bol}$ as a function of $L_{\rm bol}$. The $z \gtrsim 6$ QSOs analysed in this work occupy the high-luminosity regime and span a wide range of $\kappa_{\rm bol}^X$, with several sources lying close to the luminosity-dependent relation of \citealp{Duras2020}. Elevated bolometric corrections indicate a lower rest-frame 2--10~keV output per unit $L_{\rm bol}$, which can be naturally explained by a steeper continuum. In the context of extreme accretion, an enhanced soft-photon field from the disc can increase Compton cooling and lower the coronal temperature, leading to steep X-ray continua and, consequently, to larger $\kappa_{\rm bol}^X$ values. While uncertainties in $L_{\rm bol}$ and the non-simultaneity of UV and X-ray measurements contribute to the dispersion, the overall distribution of $\kappa_{\rm bol}^X$ supports a scenario in which a significant fraction of luminous $z \gtrsim 6$ QSOs host cold, efficiently Compton-cooled coronae.
\begin{figure}
    \centering
    \includegraphics[width=0.9\columnwidth]{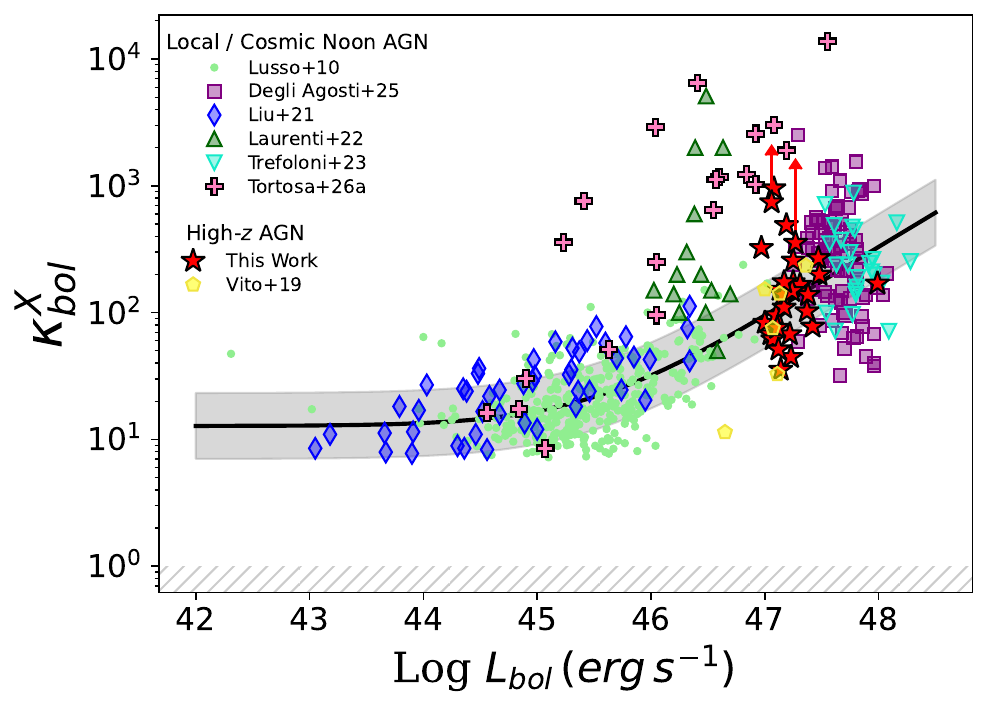}
    \caption{$\kappa^{x}_{\rm bol}$ as a function of $L_{\rm bol}$. The solid black line indicates the relation from \citealt{Duras2020}, with the shaded region showing its $1\sigma$ uncertainty. The sources analysed in this work are the red stars. We plot for comparison the following samples: X-ray selected type 1 AGN in XMM-COSMOS $0.04<z<4.5$ from \citealt{Lusso2010} (green points), WISSH QSOs at $z\sim 2-4$ from \citealt{2025A&A...702A.114D} (purple squares), local AGN/NLS1s from \citealt{2021ApJ...910..103L} (blue diamonds), nearby $0.40<z<0.75$ highly accreting AGN from \citealt{2022A&A...657A..57L} (dark green triangles), SDSS DR7 QSOs at $z\simeq 3.0-3.3$ from \citealt{2023A&A...677A.111T} (turquoise inverted triangles), $z \gtrsim 6$ QSOs from \citealt{Vito2019} (yellow pentagons), local highly accreting AGN from \citealt{Tortosa2023} (pink crosses).}
    \label{fig:kbol_Lbol}
\end{figure}
\subsection{Impact of spectral steepness on future X-ray surveys aimed at the $z \gtrsim 6$ Universe}
\label{sec:bias_steepness}
One of the main implications of our results concerns how the expected observed-band count rates of AGN at the EoR depend on their spectral shape, at fixed intrinsic luminosity, in future X-ray surveys. In the local Universe, numerous studies have shown that the X-ray photon index exhibits no significant dependence on AGN luminosity over several orders of magnitude in luminosity\citep[e.g.][]{Piconcelli2005,Ricci2017}. Our results indicate that luminous QSOs at $z\gtrsim6$ are characterised by systematically steeper X-ray continua than equally luminous QSOs at lower redshift. If the lack of a luminosity dependence of $\Gamma$ also holds in the early Universe, it is reasonable to expect that steep X-ray spectra are not restricted to the most luminous QSOs investigated here, but may characterise a substantial fraction of the AGN population at the EoR. Although this hypothesis remains to be verified observationally, it provides a physically motivated framework for evaluating how steep X-ray spectra change the expected observed-band count rates, at fixed intrinsic $L_{2-10\,\rm keV}$, of high-redshift AGN in future surveys targeting the $z \gtrsim 6$ Universe.\\
\indent Indeed, at $z > 6$, these steep X-ray continua dramatically reduce the observed-band count-rates compared to standard flatter spectra. This observational bias implies that next-generation X-ray surveys aimed at the EoR may find these early sources to be much fainter than currently expected. To quantify the observational bias introduced by steep X-ray spectra at high-$z$, we carried out a suite of simulations for current (e.g. \xmm, \textit{Chandra}) and forthcoming  (e.g. \textit{NewAthena} Wide Field Imager \citealp[WFI,][]{Cruise25}) X-ray observatories. The aim of these simulations is to assess how the observed count rate (ctr) of sources with the same luminosity varies with the intrinsic spectral shape of AGN. Specifically, we fixed the intrinsic rest-frame 2--10~keV luminosity and computed the expected count rates in the observed 0.3--2~keV, 0.3--7~keV and 2--10~keV bands. This approach isolates the effects of the spectral shape alone and removes degeneracies with intrinsic luminosity.\\
\indent We assumed a simple power-law spectrum and varied the photon index over a wide interval ($1.5 \leq \Gamma \leq 5$), exploring the redshift range $5 \leq z \leq 10$. For each pair of $(\Gamma, z)$, we calculated the corresponding expected count rate in a given energy band (EB): $ctr(\Gamma,z)_{EB}$, and defined the “spectral steepness bias” as the following normalised ratio:
\begin{equation}
 R(\Gamma)_{EB} = \frac{\mathrm{ctr}(\Gamma,z)_{EB}}{\mathrm{ctr}(\Gamma=1.9,z)_{EB}} 
\end{equation}
where $\Gamma=1.9$ is adopted as the canonical value typically used for AGN.\\
\indent We further investigated the role of the high-energy cutoff by adopting a cutoff power-law model with fixed $\Gamma=1.9$ and varying the cutoff energy $E_{\mathrm c}$ between 10 and 200~keV. In this case, we defined the “spectral cutoff energy bias”: 
\begin{equation}
R(E_{\mathrm c})_{EB} = \frac{\mathrm{ctr}(E_{\mathrm c},z)_{EB}}{\mathrm{ctr}(E_{\mathrm c}=150~\mathrm{keV},z)_{EB}}
\end{equation}
where $E_{\mathrm c}=150$~keV is chosen as a representative value of the coronal temperatures commonly inferred in AGN \citep[e.g.][]{Fabian2015,2018A&A...614A..37T,Ricci_2018}.
\begin{figure*}
    \sbox\sidefigbox{\includegraphics[width=6cm]{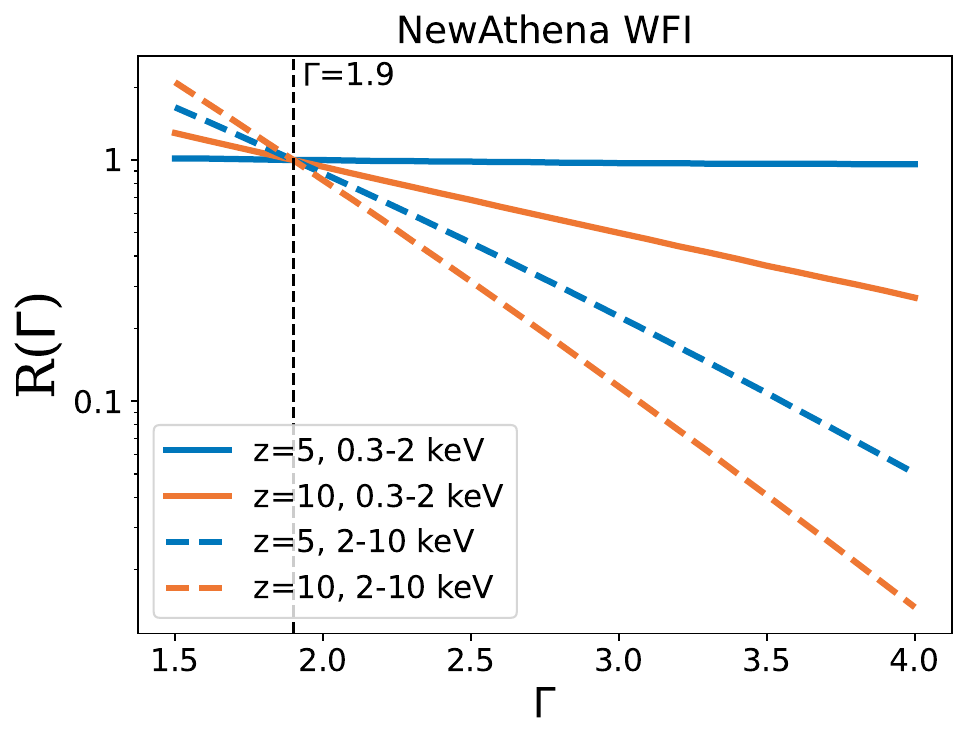}\includegraphics[width=6cm]{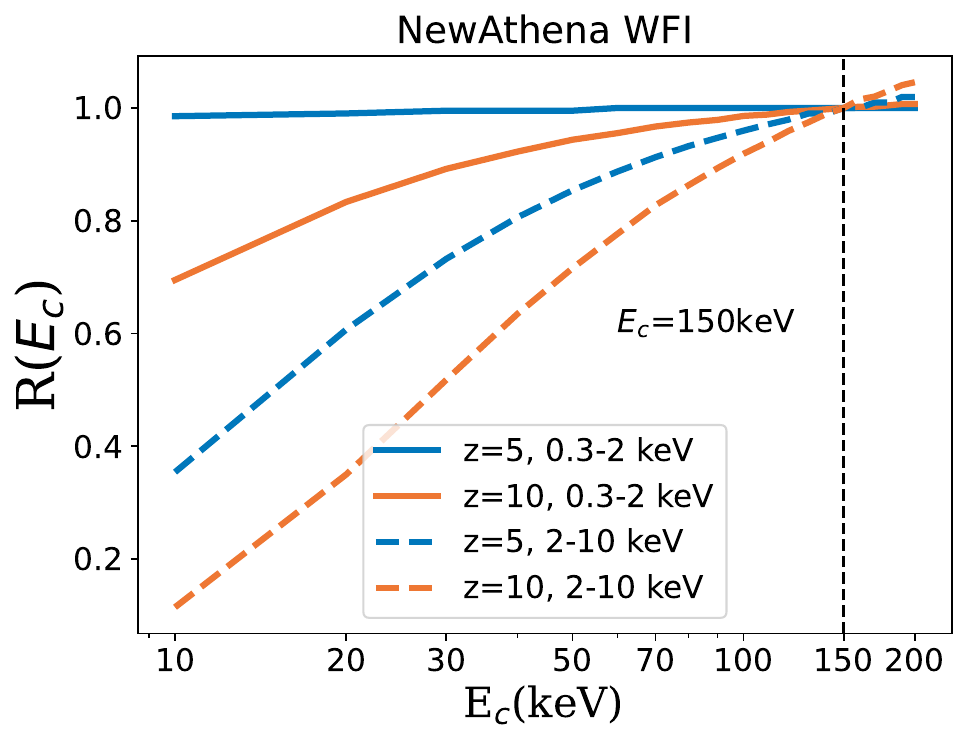}}%
    \setlength\sidecapht{\ht\sidefigbox}%
    \sidecaption
    \usebox\sidefigbox
    \caption{\textit{Left Panel:} The "spectral steepness bias" as a function of photon index ($\Gamma$) for \textit{NewAthena}/WFI; \textit{Right Panel}: The "spectral cutoff energy bias" as a function of high-energy cutoff ($E_{\rm c}$) for \textit{NewAthena}/WFI. Count rates ratios in the 0.3--2~keV (solid lines) and 2--10~keV (dashed lines) bands are shown for redshift $z=5$ (blue) and $z=10$ (orange). More detailed curves are shown in Fig.~\ref{fig:bias_gamma} and Fig.~\ref{fig:bias_Ec}}
    \label{fig:bias_NA}
\end{figure*}
Fig.~\ref{fig:bias_NA} shows how the expected \textit{NewAthena}/WFI count rates depend on the assumed X-ray spectral parameters, highlighting the potential observational biases introduced by steep continua and low cutoff energies. More detailed count-rate predictions for \textit{NewAthena}, as well as for current X-ray facilities such as \xmm\ and \textit{Chandra}, are presented Fig.~\ref{fig:bias_gamma} and Fig.~\ref{fig:bias_Ec}.\\
\indent The outcomes of our simulations highlight two distinct regimes, determined by the energy band used for count rate detection that can substantially influence the efficiency of AGN detection in X-ray surveys targeting the EoR.\\
\indent In the soft bands (0.3--2 and 0.3--7~keV), where the observed bandpass largely overlaps the rest-frame 2--10~keV normalisation band, the bias is modest for the average slope ($\langle\Gamma\rangle=2.4$, $R(\Gamma)_{EB}\approx0.8-0.9$) but becomes severe for the steepest sources ($\Gamma=3-3.5$), with count rates reduced by factors of 2--3 for \xmm/\textit{NewAthena} and 5--10 for \textit{Chandra} at $z=10$. In the harder 2--10~keV band, where the emission is dominated by extrapolation above 10~keV rest-frame, the bias is systematically stronger: factors of $\sim3$ for $\Gamma=2.4$ and up to an order of magnitude for $\Gamma=3-3.5$ ($R(\Gamma)_{EB}\approx0.04-0.1$). The analogous "cutoff energy bias" is generally milder, reaching at most $R(E_{\rm c})_{EB}\approx0.2-0.8$ for a $\sim20$~keV cutoff depending on band and redshift.\\
\indent These findings suggest that, under the hypothesis that steep spectra are a common property of the bulk of the AGN population at $z>6$, X-ray surveys targeting the EoR with future missions such as \textit{NewAthena} must account for these spectral biases when converting between intrinsic luminosity and observed flux/counts. Moreover, even if only a subset of high-$z$ AGN exhibited $\Gamma \gtrsim 2.4$, relying solely on canonical AGN spectra ($\Gamma\simeq1.8-1.9$) to predict count rates would systematically overestimate the observed-band flux of steep-spectrum sources at a given intrinsic luminosity. Translating these count-rate changes into actual survey detectability would additionally require the specific exposure time, background level, detection threshold, PSF, survey strategy, flux limit, sky coverage, and AGN luminosity function, none of which are modelled here. We also note that our calculation fixes the intrinsic $L_{2-10\,\rm keV}$. A different coupling of the disc/corona system is however possible, in which case normalising at fixed $L_{2500\,\textup{\AA}}$ or $L_{\rm bol}$ (which requires an additional assumption on $\kappa_{\rm bol}^{\rm X}$) may be at least as physically relevant.\\
\indent Similar effects may already affect current \textit{Chandra} Frontier Fields surveys aimed at investigating the X-ray emission from JWST-detected AGN \cite[e.g.][]{2024ApJ...969L..18A}. In particular, intrinsically steep X-ray continua and the associated redistribution of flux outside the observed band reduce the observed-band count rate at fixed intrinsic luminosity, which may in turn favour the selection of flatter-spectrum sources or systems characterised by high $E_{\rm c}$ values in flux-limited surveys. Although next-generation facilities will provide substantially improved sensitivity, these effects remain non-negligible even for future missions and need to be properly taken into account. In addition, proper spectral modelling and broad-band coverage will also be essential to accurately characterise the coronal emission and rapid early growth of SMBHs at the EoR.
\section{Conclusions and perspectives}
\label{sect:conclusions}
We have presented an extended X-ray spectral analysis of luminous QSOs at $z \gtrsim 6$, combining the original HYPERION sample, additional luminous non-HYPERION sources, and newly observed targets from a dedicated \xmm\ Large Programme. This expanded dataset incorporates recent high-quality SMBH mass estimates, including 10 Balmer-line-based estimates from JWST. By significantly increasing the dynamic range in mass growth rate of the SMBH powering these QSOs, while preserving comparable bolometric luminosities, this dataset enables a robust reassessment of the connection between coronal properties and early SMBH growth. Our main results can be summarised as follows:\\
$\bullet$ QSOs at the Epoch of Reionization exhibit systematically steep X-ray spectra, characterised either by high photon indices (average: $\langle\Gamma\rangle=2.4$, standard deviation of the distribution of $\Gamma$: $\sigma=0.38$) or by relatively low high-energy cutoffs when modelled with fixed $\Gamma=1.9$ (see Fig.\ref{fig:mseed_plots}). The restricted mean cutoff energy for our $z \gtrsim 6$ QSOs estimated via survival-analysis estimate is $\langle E_{\rm c}\rangle_{\rm KM}=30.0\pm 3.7$ keV (see Fig.~\ref{fig:Ec_distrib}). The cutoff energies, when measurable, occupy the low-$E_{\mathrm{c}}$, high-luminosity region of the $E_{\mathrm{c}}$–$L_{2-10\,\mathrm{keV}}$ plane (see Fig.~\ref{fig:Ec_Lx}), distinct from lower-$z$ luminous AGN.\\
$\bullet$ The persistence of steep continua at all mass growth rate regimes (see Fig.~\ref{fig:mseed_plots}) suggests that this property is not primarily driven by a selection bias in the original HYPERION sample and appear to be a common characteristic of luminous SMBHs powering luminous QSOs at the EoR. The persistence of the $\Gamma$--$v_{\rm C,\textsc{iv}}$ correlation in the expanded sample (Fig.~\ref{fig:vciv_gamma}) together with the widespread X-ray spectral steepness of luminous EoR QSOs supports the scenario of \citetalias{Tortosa2024}, in which the X-ray corona is efficiently cooled by the strong radiation field associated with high accretion states.\\
$\bullet$ The $\alpha_{\rm OX}$ and $\kappa^{x}_{\rm bol}$ of the $z \gtrsim 6$ QSOs analysed here are broadly consistent with the trends observed in highly accreting AGN at lower-$z$ (see Fig.~\ref{fig:aox_LUV} and \ref{fig:kbol_Lbol}). In particular, restricting the analysis to the HYPERION and X-LP QSOs, which were selected independently of their X-ray properties, we find an X-ray-weak fraction of $13^{+13}_{-7}\%$ (3/23, including the upper limit on J0313-1806), rising to at most $\sim17\%$ if the unclassified PSO~J183-12 is also X-ray weak (Sect.~\ref{subsec:X-rayandUV}). This fraction is lower than, but consistent within the large statistical uncertainties with, the $\sim25$--$30\%$ reported for other highly accreting QSO samples.\\
$\bullet$ Under the limiting assumption that the bulk of the EoR AGN population shares the steep $\Gamma$ measured here, future X-ray surveys (e.g. with \textit{NewAthena}) could face observational biases translating into expected count-rate reductions, at fixed intrinsic $L_{2-10\,\rm keV}$, with consequences for the detectability and census of early SMBH growth. For the sample average ($\langle\Gamma\rangle\sim2.4$), this reduction is modest ($\lesssim20\%$) in the soft bands most relevant for detecting faint high-$z$ sources, while it can reach factors of 10 or more for the steepest spectra in our sample ($\Gamma\sim3$--3.5) and/or in the hard 2--10~keV observed band. This bounding-case estimate, while not a forecast of the full EoR population, should be factored into the planning of future high-$z$ X-ray surveys.\\
$\bullet$ Steep X-ray continua appear to be a widespread property of luminous QSOs at the EoR and are closely linked to strong disc winds, but do not directly trace the integrated seed-mass growth history. Future high-sensitivity, broad-band X-ray missions such as \textit{NewAthena} \citep{Cruise25} will be crucial to directly measure the coronal temperatures and disentangle spectral slope-cutoff energy degeneracies, thereby placing definitive constraints on the physical state of the accretion flow in the first generation of SMBHs. In addition, their improved sensitivity will enable the characterisation of lower-luminosity AGN at high-$z$, allowing us to test whether steep X-ray spectra are a ubiquitous property of early SMBH growth or are instead confined to the most luminous and rapidly accreting systems.\\
\indent The findings reported here also strongly motivate investing very long exposures of current facilities, such as \xmm, on carefully selected individual $z \gtrsim 6$ QSOs. Such data could place key constraints on the coronal emission of EoR AGN, enabling us to distinguish between intrinsically steep continua and low high-energy cutoffs, while simultaneously building the spectral templates required to properly interpret JWST-selected AGN and to optimise future high-$z$ X-ray survey strategies.
\begin{acknowledgements}
    This work is based on observations obtained with the ESA science mission {\it XMM-Newton}, with instruments and contributions directly funded by ESA Member States and the USA (NASA). 
    ChatGPT (OpenAI) was used for language editing and code debugging during the preparation of this manuscript.
    The authors acknowledge support from the INAF Large Program “DELUX” of the “Ricerca Fondamentale 2024” INAF program and from the Bando Ricerca Fondamentale INAF 2022 Large Grant “Toward an holistic view of the Titans: multi-band observations of $z \gtrsim 6$ QSOs powered by greedy supermassive black holes".
    EB acknowledges funding from the Bando Ricerca Fondamentale INAF 2024 GO grant "A JWST/MIRI MIRACLE: Mid-IR Activity of Circumnuclear Line Emission" and MiniGrant RSN1 1.05.24.07.01.
    FS and CF acknowledge financial support from the Ricerca Fondamentale INAF 2023 Data Analysis grant 1.05.23.03.04 ``ARCHIE ARchive Cosmic HI \& ISM  Evolution'', Ricerca Fondamentale INAF 2024 under project 1.05.24.07.01 MINI-GRANTS RSN1 "ECHOS", Bando Finanziamento ASI CI-UCO-DSR-2022-43 CUP:C93C25004260005 project ``IBISCO: feedback and obscuration in local AGN''.
    GM acknowledges support from grant PID2023-147338NB-C21 funded by MICIU/AEI/10.13039/501100011033 and ERDF/EU.
    MC acknowledges INAF GO Grant 2024 ”Revealing the nature of bright galaxies at cosmic dawn with deep JWST spectroscopy”.
    MP acknowledges support through the grants PID2021-127718NB-I00, PID2024-159902NA-I00, and RYC2023-044853-I, funded by the Spain Ministry of Science and Innovation/State Agency of Research MCIN/AEI/10.13039/501100011033 and El Fondo Social Europeo Plus FSE+. 
    VD and CP acknowledge support from the ERC synergy grant 101166930 - RECAP.
\end{acknowledgements}

\bibliographystyle{aa}
\bibliography{biblio}
\begin{appendix}
\section{Black hole masses}
\label{app:masses}
To ensure a homogeneous determination of the black hole mass across the full sample, we re-derived all virial mass estimates using consistent scaling relations and cosmological assumptions.

For sources with Mg~\textsc{ii}-based measurements available in the literature, we collected the reported continuum luminosities at 3000~\AA\ and Mg~\textsc{ii} full width at half maximum (FWHM) values, and recomputed the black hole masses using the calibration of \citealt{2009ApJ...699..800V}. Continuum luminosities were re-evaluated adopting the same cosmology used throughout this work, consistent with that assumed in \textsc{xspec}, in order to avoid systematic offsets due to heterogeneous assumptions in the literature.

For sources with rest-frame optical spectroscopy from \textit{JWST}, black hole masses were derived from the broad H$\alpha$ and H$\beta$ emission lines. In particular, H$\alpha$-based masses were computed following the scaling relation of \citealt{2011ApJS..194...45S}, while H$\beta$-based masses were obtained using the relation of \citealt{2006ApJ...641..689V}. In all cases, line and continuum luminosities were computed adopting the same cosmological parameters, ensuring internal consistency among the different mass estimators.

We find that the typical difference between masses derived using different estimators is large, with an average spread of $\langle \Delta \log M_{\rm BH} \rangle \simeq 0.3$ dex. This intrinsic scatter between different virial estimators largely dominates over the differences introduced by adopting alternative calibrations for the same emission line. As a consequence, the specific choice of calibration within a given estimator (e.g. different H$\beta$-based relations) has a negligible impact on our results compared to the systematic differences between distinct estimators.

From the black hole masses derived as described above, we also estimated the \mseed\ parameter assuming continuous accretion at the Eddington limit ($\lambda_{\rm Edd}=1$), following the exponential growth relation: $M_{\rm s,Edd} = M_{\rm BH} \times e^{-t/t_{\rm s}}$ where $t$ is the time elapsed between the formation redshift, $z_{\rm f}$, of the seed black hole and the redshift at which the QSO is observed, and $t_{\rm s}$ is the e-folding time, $t_{\rm s} = 0.45\,\eta\,(1-\eta)^{-1}\,\lambda_{\rm Edd}^{-1}\,f_{\rm duty}^{-1}\ {\rm Gyr}$. In this calculation, we assumed $z_{\rm f} = 20$ \citep{2016MNRAS.457.3356V}, a radiative efficiency $\eta = 0.1$, representing the fraction of accreted mass converted into radiation, and a duty cycle $f_{\rm duty} = 1$, corresponding to an uninterrupted active phase.

All the values of the black hole masses and \mseed\ we estimated are reported in Tab.~\ref{table:masses} and graphically showed in Fig.~\ref{fig:masses}

\begin{table*}\caption{$M_{\rm BH}$ and \mseed\ for our sources, estimated using consistent scaling relations and cosmological assumptions (see App.~\ref{app:masses}). \mseed\ values have been recomputed from $M_{\rm s,Edd} = M_{\rm BH}\,e^{-t/t_s}$ for internal consistency.}
\label{table:masses} 
\centering 
\small
\begin{tabular}{lrrrrll}
\hline\hline
  \multicolumn{1}{c}{Target } &
  \multicolumn{1}{c}{$M_{\rm BH}/M_{\odot} [10^8]$} &
  \multicolumn{1}{c}{$<M_{\rm BH}>$} &
  \multicolumn{1}{c}{\mseed/$M_{\odot} [10^3]$} &
    \multicolumn{1}{c}{<\mseed>} &
  \multicolumn{1}{l}{Ref.} &
  \multicolumn{1}{l}{Method} \\
\hline
  ULAS~J1342+0928 & 8.89 $\pm$ 3.41 & \multirow{4}{*}{$10.00^{+5.89}_{-4.52}$} & 21.61 $\pm$ 8.29 & \multirow{4}{*}{$24.31^{+14.33}_{-11.00}$} & \citealt{2018Natur.553..473B} & Mg~\textsc{ii}\\
   & 9.22 $\pm$ 1.37 &  & 22.41 $\pm$ 3.33 &  & \citealt{Onoue2020} & Mg~\textsc{ii}\\
   & 8.41 $\pm$ 2.26 &  & 20.44 $\pm$ 5.49 &  & \citealt{Yang2021} & Mg~\textsc{ii}\\
   & 13.50 $\pm$ 2.40 &  & 32.81 $\pm$ 5.83 &  & \citealt{Trefoloni2026} & Mg~\textsc{ii}\\

   & 4.05 $\pm$ 0.42 &  & 9.84 $\pm$ 1.02 &  & \citealt{2025arXiv251102902B} & H$\alpha$\\

   & 11.30 $\pm$ 3.10 &  & 27.46 $\pm$ 7.53 &  & \citealt{Trefolono2026} & H$\beta$\\
  \hline
  J1007+2115 & 14.80 $\pm$ 3.06 & \multirow{2}{*}{$14.95^{+2.91}_{-3.21}$} & 31.93 $\pm$ 6.60 & \multirow{2}{*}{$32.25^{+6.28}_{-6.93}$} & \citealt{Yang2021} & Mg~\textsc{ii}\\
   & 15.10 $\pm$ 1.79 &  & 32.58 $\pm$ 3.86 &  & \citealt{2020ApJ...897L..14Y} & Mg~\textsc{ii}\\

   & 7.90 $\pm$ 0.85 &  & 17.04 $\pm$ 1.83 &  & \citealt{2025arXiv251102902B} & H$\alpha$\\
  \hline
  ULAS~J1120+0641 & 25.80 $\pm$ 6.70 & \multirow{3}{*}{$21.20^{+11.30}_{-10.49}$} & 18.45 $\pm$ 4.79 & \multirow{3}{*}{$15.16^{+8.08}_{-7.50}$} & \citealt{2017ApJ...849...91M} & Mg~\textsc{ii}\\
   & 14.00 $\pm$ 3.29 &  & 10.01 $\pm$ 2.35 &  & \citealt{Yang2021} & Mg~\textsc{ii}\\
   & 23.80 $\pm$ 2.58 &  & 17.02 $\pm$ 1.85 &  & \citealt{2014ApJ...790..145D} & Mg~\textsc{ii}\\

   & 14.80 $\pm$ 1.21 &  & 10.59 $\pm$ 0.87 &  & \citealt{2025arXiv251102902B} & H$\alpha$\\
  \hline
  DELS~J0038-1527 & 13.80 $\pm$ 1.53 & \multirow{2}{*}{$13.95\pm1.68$} & 8.14 $\pm$ 0.90 & \multirow{2}{*}{$8.23\pm0.99$} & \citealt{Wang2018} & Mg~\textsc{ii}\\
   & 14.10 $\pm$ 1.53 &  & 8.32 $\pm$ 0.90 &  & \citealt{Yang2021} & Mg~\textsc{ii}\\
  \hline
  DES~J0252-0503 & 14.30 $\pm$ 1.76 & \multirow{2}{*}{$13.80^{+2.26}_{-1.95}$} & 7.70 $\pm$ 0.95 & \multirow{2}{*}{$7.43^{+1.22}_{-1.05}$} & \citealt{2020ApJ...896...23W} & Mg~\textsc{ii}\\
   & 13.30 $\pm$ 1.45 &  & 7.16 $\pm$ 0.78 &  & \citealt{Yang2021} & Mg~\textsc{ii}\\
  \hline
  DES~J0020-3653 & 17.60 $\pm$ 3.33 &  & 5.89 $\pm$ 1.11 &  & \citealt{Reed2019} & Mg~\textsc{ii}\\

   & 11.90 $\pm$ 0.32 &  & 3.98 $\pm$ 0.11 &  & \citealt{marshall2023} & H$\alpha$\\

   & 26.10 $\pm$ 2.72 &  & 8.74 $\pm$ 0.91 &  & \citealt{marshall2023} & H$\beta$\\
  \hline
  VHS~J0411-0907 & 6.36 $\pm$ 0.53 & \multirow{2}{*}{$8.10^{+3.62}_{-2.27}$} & 2.06 $\pm$ 0.17 & \multirow{2}{*}{$2.63^{+1.18}_{-0.74}$} & \citealt{2019MNRAS.484.5142P} & Mg~\textsc{ii}\\
   & 9.85 $\pm$ 1.88 &  & 3.20 $\pm$ 0.61 &  & \citealt{Yang2021} & Mg~\textsc{ii}\\

   & 9.59 $\pm$ 0.80 &  & 3.11 $\pm$ 0.26 &  & \citealt{marshall2023} & H$\alpha$\\

   & 16.90 $\pm$ 1.51 &  & 5.48 $\pm$ 0.49 &  & \citealt{marshall2023} & H$\beta$\\
  \hline
  VDES~J0244-5008 & 12.10 $\pm$ 4.12 &  & 2.86 $\pm$ 0.97 &  & \citealt{Reed2019} & Mg~\textsc{ii}\\

   & 8.63 $\pm$ 0.91 &  & 2.04 $\pm$ 0.21 &  & \citealt{2023ApJ...951L...5Y} & H$\beta$\\
  \hline
  PSO~J231.6-20.8 & 42.20 $\pm$ 8.03 & \multirow{2}{*}{$38.20^{+12.03}_{-6.66}$} & 6.33 $\pm$ 1.21 & \multirow{2}{*}{$5.73^{+1.81}_{-1.00}$} & \citealt{Farina2022} & Mg~\textsc{ii}\\
   & 34.20 $\pm$ 2.66 &  & 5.13 $\pm$ 0.40 &  & \citealt{Mazzucchelli2023} & Mg~\textsc{ii}\\
  \hline
  PSO~J036.5+03.0 & 31.20 $\pm$ 6.28 & \multirow{2}{*}{$34.95^{+9.21}_{-10.03}$} & 3.89 $\pm$ 0.78 & \multirow{2}{*}{$4.36^{+1.15}_{-1.25}$} & \citealt{2017ApJ...849...91M} & Mg~\textsc{ii}\\
   & 38.70 $\pm$ 5.46 &  & 4.83 $\pm$ 0.68 &  & \citealt{Farina2022} & Mg~\textsc{ii}\\
  \hline
  VDES~J0224-4711 & 12.40 $\pm$ 0.19 & \multirow{2}{*}{$17.55^{+9.17}_{-5.34}$} & 1.51 $\pm$ 0.02 & \multirow{2}{*}{$2.14^{+1.12}_{-0.65}$} & \citealt{Mazzucchelli2023} & Mg~\textsc{ii}\\
   & 22.70 $\pm$ 4.02 &  & 2.77 $\pm$ 0.49 &  & \citealt{Reed2019} & Mg~\textsc{ii}\\

   & 22.30 $\pm$ 0.83 &  & 2.72 $\pm$ 0.10 &  & \citealt{2023ApJ...951L...5Y} & H$\beta$\\
  \hline
  PSO~J011+09 & 13.80 $\pm$ 4.68 & \multirow{2}{*}{$9.28^{+9.20}_{-6.11}$} & 1.26 $\pm$ 0.43 & \multirow{2}{*}{$0.85^{+0.84}_{-0.56}$} & \citealt{Eilers2020} & Mg~\textsc{ii}\\
   & 4.76 $\pm$ 1.59 &  & 0.44 $\pm$ 0.15 &  & \citealt{Farina2022} & Mg~\textsc{ii}\\
  \hline
  SDSS~J1148+5251 & 32.40 $\pm$ 2.05 & \multirow{2}{*}{$43.30^{+13.26}_{-12.95}$} & 2.74 $\pm$ 0.17 & \multirow{2}{*}{$3.66^{+1.12}_{-1.10}$} & \citealt{Farina2022} & Mg~\textsc{ii}\\
   & 54.20 $\pm$ 2.36 &  & 4.59 $\pm$ 0.20 &  & \citealt{Shen2019} & Mg~\textsc{ii}\\
  \hline
  PSO~J083.8+11.8 & 20.70 $\pm$ 4.30 &  & 1.33 $\pm$ 0.28 &  & \citealt{2020ApJ...903...34A} & Mg~\textsc{ii}\\
  \hline
  SDSS~J0100+2802 & 153.00 $\pm$ 9.04 & \multirow{4}{*}{$106.03^{+56.01}_{-51.08}$} & 8.30 $\pm$ 0.49 & \multirow{4}{*}{$5.75^{+3.04}_{-2.77}$} & \citealt{Mazzucchelli2023} & Mg~\textsc{ii}\\
   & 114.00 $\pm$ 0.08 &  & 6.18 $\pm$ 0.00 &  & \citealt{Wu2015} & Mg~\textsc{ii}\\
   & 101.00 $\pm$ 2.52 &  & 5.48 $\pm$ 0.14 &  & \citealt{Farina2022} & Mg~\textsc{ii}\\
   & 56.10 $\pm$ 1.16 &  & 3.04 $\pm$ 0.06 &  & \citealt{Eilers_2023} & Mg~\textsc{ii}\\

   & 166.00 $\pm$ 8.44 &  & 9.00 $\pm$ 0.46 &  & \citealt{Eilers_2023} & H$\beta$\\
  \hline
  ATLAS~J025-33 & 24.40 $\pm$ 11.40 & \multirow{2}{*}{$25.50^{+10.30}_{-12.50}$} & 1.29 $\pm$ 0.60 & \multirow{2}{*}{$1.35^{+0.55}_{-0.66}$} & \citealt{Mazzucchelli2023} & Mg~\textsc{ii}\\
   & 26.60 $\pm$ 7.90 &  & 1.41 $\pm$ 0.42 &  & \citealt{2018MNRAS.478.1649C} & Mg~\textsc{ii}\\
  \hline
  CFHQS~J0050+3445 & 47.90 $\pm$ 8.02 &  & 2.12 $\pm$ 0.36 &  & \citealt{2011ApJS..194...45S} & Mg~\textsc{ii}\\
  \hline
  ATLAS~J029-36 & 16.60 $\pm$ 2.42 & \multirow{2}{*}{$41.95^{+40.65}_{-27.77}$} & 0.31 $\pm$ 0.05 & \multirow{2}{*}{$0.78^{+0.76}_{-0.52}$} & \citealt{Mazzucchelli2023} & Mg~\textsc{ii}\\
   & 67.30 $\pm$ 15.30 &  & 1.26 $\pm$ 0.29 &  & \citealt{2018MNRAS.478.1649C} & Mg~\textsc{ii}\\
  \hline
  PSO~J159-02 & 32.50 $\pm$ 3.42 &  & 2.36 $\pm$ 0.25 &  & \citealt{Farina2022} & Mg~\textsc{ii}\\
  \hline
  SDSS~J1030+0524 & 19.40 $\pm$ 4.32 & \multirow{3}{*}{$18.43^{+5.28}_{-4.04}$} & 1.08 $\pm$ 0.24 & \multirow{3}{*}{$1.03^{+0.30}_{-0.23}$} & \citealt{Mazzucchelli2023} & Mg~\textsc{ii}\\
   & 20.00 $\pm$ 2.19 &  & 1.12 $\pm$ 0.12 &  & \citealt{Farina2022} & Mg~\textsc{ii}\\
   & 15.90 $\pm$ 1.51 &  & 0.89 $\pm$ 0.08 &  & \citealt{2011ApJS..194...45S} & Mg~\textsc{ii}\\
  \hline
  SDSS~J1602+4228 & 15.90 $\pm$ 1.51 &  & 0.38 $\pm$ 0.04 &  & \citealt{2011ApJS..194...45S} & Mg~\textsc{ii}\\
  \hline
  PSO~J308-21 & 17.40 $\pm$ 3.75 &  & 0.42 $\pm$ 0.09 &  & \citealt{Farina2022} & Mg~\textsc{ii}\\

   & 38.10 $\pm$ 1.19 &  & 0.92 $\pm$ 0.03 &  & \citealt{Loiacono2024} & H$\beta$\\
  \hline
  SDSS~J1306+0356 & 21.70 $\pm$ 0.65 &  & 0.42 $\pm$ 0.01 &  & \citealt{Farina2022} & Mg~\textsc{ii}\\
  \hline
  J0313-1806 & 16.60 $\pm$ 3.67 &  & 51.57 $\pm$ 11.40 &  & \citealt{Wang_2021} & Mg~\textsc{ii}\\

   & 7.59 $\pm$ 0.65 &  & 23.58 $\pm$ 2.02 &  & \citealt{2025arXiv251102902B} & H$\alpha$\\
  \hline
  SDSS~J2054-0005 & 15.30 $\pm$ 1.95 &  & 0.27 $\pm$ 0.03 &  & \citealt{Farina2022} & Mg~\textsc{ii}\\
  \hline
  SDSS~J1411+1217 & 8.48 $\pm$ 1.94 &  & 0.10 $\pm$ 0.02 &  & \citealt{Jiang_2007} & Mg~\textsc{ii}\\
  \hline
  PSO~J183-12 & 17.20 $\pm$ 5.71 &  & 0.16 $\pm$ 0.05 &  & \citealt{Mazzucchelli2023} & Mg~\textsc{ii}\\
  \hline
  PSO~J025-11 & 22.60 $\pm$ 3.37 &  & 0.20 $\pm$ 0.03 &  & \citealt{Mazzucchelli2023} & Mg~\textsc{ii}\\
  \hline
\end{tabular}
\end{table*}
\begin{figure*}
    \sbox\sidefigbox{\includegraphics[width=12cm]{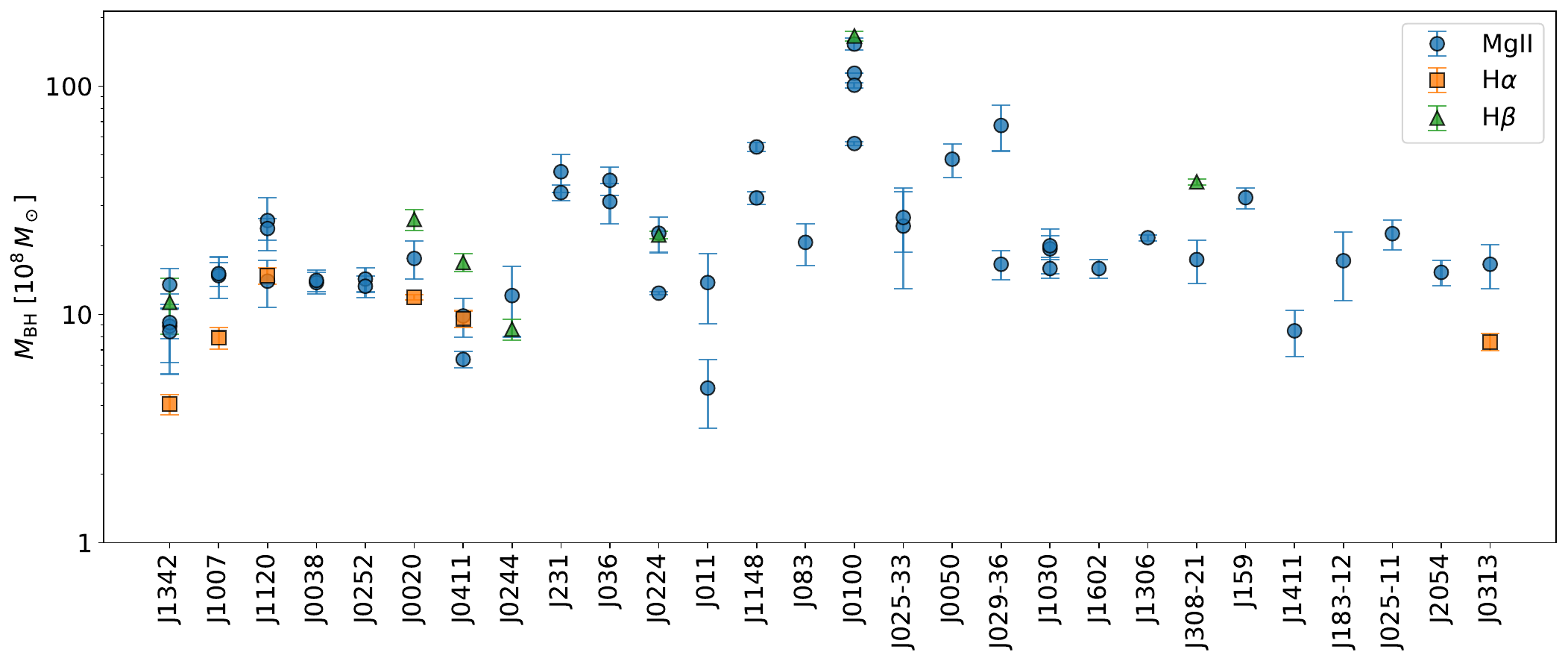}}%
    \setlength\sidecapht{\ht\sidefigbox}%
    \sidecaption
    \usebox\sidefigbox
    \caption{Black hole mass estimates, computing using consistent scaling relations and cosmological assumptions (see App.~\ref{app:masses}) for the sources listed in Tab.~\ref{table:masses}. For each object, we report the single-epoch virial masses derived from different broad emission lines: Mg~\textsc{ii} (blue circles), H$\alpha$ (orange squares), and H$\beta$ (green triangles), when available.}
    \label{fig:masses}
\end{figure*}

\FloatBarrier
\section{Data reduction}
\label{app:data_red}
\subsection{XMM-Newton}
\label{app:xmm_data_red}
We report here the details of the \xmm\ observations from the new LP in Table \ref{table:obsid}, as well as the HYPERION observations that were not been included in \citetalias{Tortosa2024}. We remand to \citetalias{Tortosa2024} for the details of the other HYPERION and non-HYPERION QSOs considered in this work.
\begin{table*}
\caption{\xmm\ observations analysed in this work including the new observations from the QSOs belonging to the recently approved X-LP and the HYPERION QSOs observation from the multi-year X-HP \citepalias{Zappacosta2023} which were not included in \citetalias{Tortosa2024}.}
\label{table:obsid}  
\centering 
\small
\begin{tabular}{llcccrrrr}   
\hline\hline            
OBSID      & Target          & RA          &DEC          &Start date                &Exp. & \multicolumn{3} {c}{Net Exp.}\\
           &                 &(J2000)      & (J2000)     & (UTC)                    & [ks]& PN [ks] & MOS1 [ks] & MOS2 [ks]  \\
\hline

\noalign{\smallskip}
\multicolumn{9} {c}{HYPERION QSOs} \\
\noalign{\smallskip}
0884994101& PSO~J036.5+03.0& 02:26:01.88& +03:02:59.4& 2024-01-09& 83.0&51.96&80.95&80.96 \\
0886202001& DELS~J0038-1527& 00:38:36.09 & -15:27:23.6& 2023-12-14&83.9& 47.58&58.03&57.02\\
0884990701& DELS~J0038-1527& 00:38:36.09 & -15:27:23.6& 2023-12-10&95.3& 63.05&85.47&85.46\\	
0884992401& PSO~J083.8+11.8&05:35:20.89& +11:50:53.6& 2022-03-13& 90.3& 52.69 &	85.44 &85.47\\
\hline
\noalign{\smallskip}
\multicolumn{9} {c}{X-LP} \\[0.1cm]
\noalign{\smallskip}
0943771101&PSO~J183-12& 12:13:11.81  & -12:46:03.4& 2025-12-15& 18 &2.11& 5.20&5.03\\
0943770201&PSO~J183-12& 12:13:11.81  & -12:46:03.4& 2025-12-15& 126.3& 27.16 & 40.24 & 40.17 \\
0943771001& PSO~J025-11& 01:40:57.02 & -11:40:59.4& 2025-07-04& 46.1 & 22.79 & 27.63 & 25.12\\
0943770901& PSO~J025-11& 01:40:57.02 & -11:40:59.4& 2025-07-03& 50.9& 30.05 & 42.14 &46.24\\
0943770301& PSO~J025-11& 01:40:57.02 & -11:40:59.4& 2025-07-02& 46.5& 24.25 & 34.58 & 34.81\\
0943770101 & SDSS~J1411+1217& 14:11:11.29& +12:17:37.4& 2025-06-28& 121.9 & 50.90 & 71.86 & 70.97\\
0943770501& J0313-1806& 03:13:43.83& -18:06:36.4& 2025-01-23& 100.1& 49.22 & 78.46 & 77.95\\
0943770801& SDSS~J2054-0005& 20:54:06.48& -00:05:14.8& 2024-10-20& 54.0& 35.92 & 46.07 & 47.36\\
0943770701& SDSS~J2054-0005& 20:54:06.48& -00:05:14.8& 2024-10-19& 61.6& 24.84 & 20.02 & 15.39\\
0943770601& J0313-1806& 03:13:43.83& -18:06:36.4& 2024-07-10& 100.7& 44.50 & 41.98 & 39.96\\
\hline\hline
\end{tabular}
\end{table*}

The data reduction of the above mentioned observations was carried out following the methodology described in \citetalias{Zappacosta2023} and \citetalias{Tortosa2024}. Event lists for the EPIC-PN and EPIC-MOS detectors were produced using the \texttt{epproc} and \texttt{emproc} tasks within the standard System Analysis Software (\texttt{SAS} v.18.0.0). We used the most recent calibration files available as of July 2025. 

To identify and exclude intervals affected by flaring particle background, we visually inspected light curves extracted in the 10–12\,keV band for EPIC-PN and above 10\,keV for EPIC-MOS, using PATTERN=0 (single-pixel events). To define the source and background extraction regions, we first detected point-like sources in each field of view by running the meta-task \texttt{edetect-chain} on the 0.5–2\,keV EPIC images, adopting a detection maximum likelihood threshold of DETML=6 (where DETML=$-\ln P_{\rm rnd}$ and $P_{\rm rnd}$ is the probability of a spurious detection). 

Source photons were extracted from circular regions with a radius of 20 arcsec (enclosing $\sim 80\%$ of the on-axis PSF encircled energy fraction at 1.5\,keV), centered on the published optical coordinates of each target. For J0313, which has two neighbouring sources, which are not associated with any catalogued multi-wavelength counterparts in the currently available ancillary data, we reduced the target extraction radius to 10 arcsec. For the EPIC-PN detector, background spectra were extracted from rectangular regions with long and short sides between 3.6–3.9 arcmin and 1.9–2.7 arcmin, respectively, positioned around the source region and aligned with the detector position angle. For each of the two EPIC-MOS cameras, background regions were chosen as circular areas with radii spanning 2.5–3.4 arcmin, centered on the target. From all background regions, we excluded circular areas of radius 40 arcsec centered on the target position and on all contaminating point sources identified in the previous step. 

Response matrices and ancillary response files were produced with the \texttt{SAS} tasks \texttt{rmfgen} and \texttt{arfgen}, respectively. The spectra were grouped using the optimal scheme of \citealt{2016A&A...587A.151K} (hereafter KB), which provides an optimal binning scheme by accounting for the intrinsic spectral shape, the energy-dependent resolution, and the mean photon energy in each bin. The KB grouping method is particularly suitable for obtaining unbiased, energy-independent spectral parameter estimates in the low-count regime typical of $z \gtrsim 6$ sources (see App. B of \citetalias{Zappacosta2023} for additional details and comparison with other grouping schemes).

\subsection{Chandra}
\label{app:Chandra_data_red}
For the QSOs observed with \textit{Chandra}, we reduced and analysed the available archival ACIS-S data using the \textit{Chandra} Interactive Analysis of Observations (\texttt{CIAO}) software v.4.16.0 and the most recent calibration files. In particular, we processed the observations of SDSS~J1602+4228 (OBSID~5609), SDSS~J1306+0356 (OBSID~3966) and PSO~J308$-$21 (OBSID~20470, 21725, and 21726). The source and background spectra were extracted using the \texttt{specextract} task. Source spectra were extracted from circular regions centered on the optical positions of the QSOs, which are consistent with the detected X-ray source positions within the astrometry uncertainties, with radii of $\sim$2~arcsec, corresponding to $\gtrsim$90\% of the encircled energy fraction at 1.5~keV for on-axis ACIS-S observations. Background spectra were extracted from nearby source-free annular regions centered on the source position, with inner and outer radii of $\sim$5~arcsec and $\sim$20~arcsec, respectively, avoiding contamination from other point sources and chip edges. For PSO~J308$-$21, which has multiple \textit{Chandra} observations, the spectra and corresponding response files were extracted separately for each OBSID and subsequently combined for the spectral analysis with the \texttt{combinespectra} task.

We note that SDSS~J1030+0524 has been observed also with a deep ($\sim 500$~ks)  ACIS-I \textit{Chandra} campaign. The \textit{Chandra} data caught the source in a significantly fainter X-ray state, likely associated with a transient dimming event \citep{2018A&A...614A.121N}, whereas the \textit{XMM} observation is more representative of the source in its bright state. Since our analysis aims at comparing the intrinsic coronal properties of luminous $z > 6$ QSOs, we consider the higher-flux \textit{XMM} observation to be more representative of the typical source state. In addition, the \textit{XMM} spectrum provides tighter constraints on $\Gamma$ despite the shorter exposure \citep{2018A&A...614A.121N}.

\FloatBarrier
\section{Additional correlation tests}
\label{app:other_relations}
For completeness, to provide the full set of tested relations, we report here additional correlation tests between the X-ray and broadband properties, \mseed\ and the $v_{\rm C~\textsc{iv}}$. The corresponding best-fit parameters, Pearson coefficients, null-hypothesis probabilities, and statistical significances are listed in Tab.~\ref{table:all_fitting_param}. In addition, we report the $v_{\rm C~\textsc{iv}}$ values for the targets of the new X-LP which were not reported in \citetalias{Tortosa2024} and for the HYPERION sources VHS~J0411-0907 and DELS~J0038-1527 which were not included in Table 3 of \citetalias{Tortosa2024}.

\begin{table*}
\caption{Best-fit linear relations (y=Ax+B) tested in this work, together with their correlation coefficients, null-hypothesis probabilities of the correlation and statistical significance. The fitting procedure did not include the sources considered as undetected (see Sect.~\ref{sect:sample_and_data_red}).}
\label{table:all_fitting_param}  
\centering 
\small
\begin{tabular}{lclccrcr}   
\hline\hline                
\multicolumn{3}{l}{Relation (y vs. x)}& Slope (A) &Intercept (B) &$\rho_{P}$ & 1-P$_{\rm null}$ & $\sigma$\\[0.1cm]
\hline\hline    
$\Gamma$                   & vs.& \mseed            & $0.05 \pm 0.11$  & $2.09 \pm 0.38$   & $0.15$  & $0.564$ & $<1$\\[0.1cm]
$E_{\rm c}$                & vs.& \mseed            & $-2.34 \pm 6.50$ & $31.61\pm 23.46$ & $-0.24$ & $0.698$ & $1.0$ \\[0.1cm]
$\alpha_{\rm OX}$          & vs.& \mseed            & $-0.08 \pm 0.10$  & $-1.38 \pm 0.37$  & $0.02$  & $0.271$ & $<1$ \\[0.1cm]
$\kappa_{\rm bol}^{\rm X}$ & vs.& \mseed            & $-0.13 \pm 0.13$  & $2.49 \pm 0.43$   & $-0.12$ & $0.556$ & $<1$\\[0.1cm]
$L_{\rm 2-10 keV}$         & vs.& \mseed            & $0.13 \pm 0.24$  & $44.84 \pm 0.75$  & $0.24$  & $0.740$ & $1.1$\\[0.1cm]
\hline
$\Gamma$                   & vs.&$v_{\rm C~\textsc{iv}}$ & $(-1.09 \pm 0.41)\times 10^{-4}$ & $2.12 \pm 0.09$ & $-0.630$ & $0.999$ & $3.3$ \\[0.1cm]
$E_{\rm c}$                & vs.&$v_{\rm C~\textsc{iv}}$ & $(2.8 \pm 4.1)\times 10^{-3}$ & $21.41 \pm 10.78$ & $0.107$ & $0.390$ & $<1$ \\[0.1cm]
$\alpha_{\rm OX}$          & vs.&$v_{\rm C~\textsc{iv}}$ & $(3.3 \pm 3.9)\times 10^{-5}$ & $-1.58 \pm 0.15$ & $0.044$ & $0.336$ & $<1$ \\[0.1cm]
$\kappa_{\rm bol}^{\rm X}$ & vs.&$v_{\rm C~\textsc{iv}}$ & $(-6.3 \pm 5.2)\times 10^{-5}$ & $1.92 \pm 0.15$ & $-0.062$ & $0.459$ & $<1$ \\[0.1cm]
$L_{\rm 2-10 keV}$         & vs.&$v_{\rm C~\textsc{iv}}$ & $(2.4 \pm 3.2)\times 10^{-5}$ & $45.32 \pm 0.13$ & $0.111$ & $0.398$ & $<1$ \\[0.1cm]
\hline
$\Gamma$                   & vs.& $\kappa_{\rm bol}^{\rm X}$& $0.14 \pm 0.16$ & $2.04 \pm 0.33$  &$0.12$& $0.537$& $<1$\\[0.1cm]

\hline
\hline
\end{tabular}
\end{table*}

\begin{figure*}
    \centering
    \includegraphics[width=1.\textwidth]{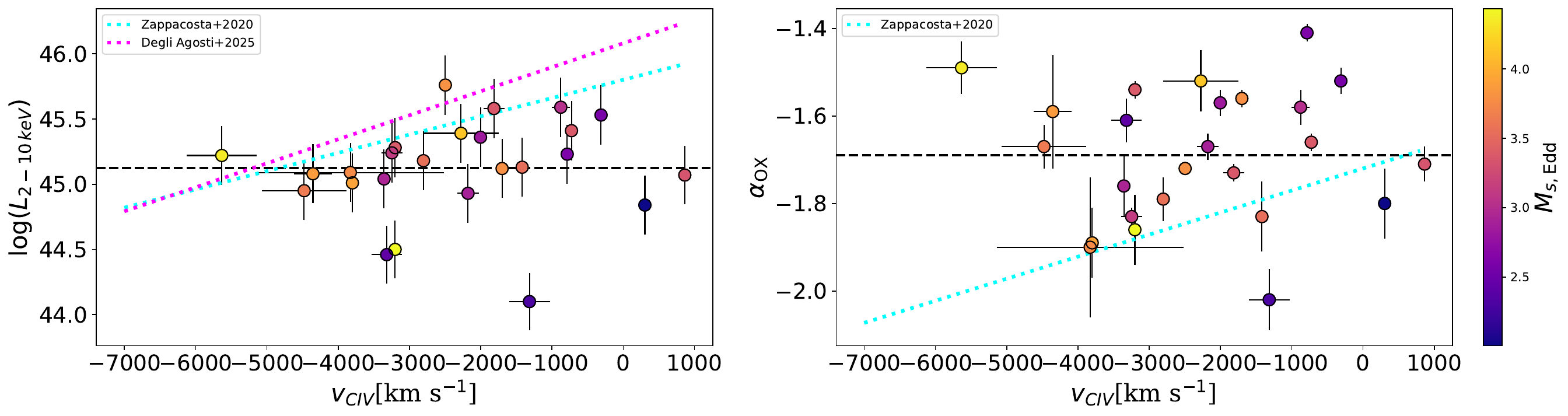}
    \caption{\textit{Left panel}: $L_{\rm 2-10 keV}$ as a function of $v_{\rm C~\textsc{iv}}$; \textit{Right panel}: $\alpha_{\rm OX}$ as a function of $v_{\rm C~\textsc{iv}}$. Points are colour coded for \mseed. The horizontal dashed line in each panel represents the mean value of the corresponding y-parameter. The cyan and magenta dotted lines represent the relations derived for the WISSH QSOs sample by \citet{Zappacosta2020} and \citet{2025A&A...702A.114D}, respectively.}
    \label{fig:vciv_plots}
\end{figure*}

\begin{table}
\caption{Values of $v_{\rm C\,\textsc{iv}}$ and $REW_{\rm C\,\textsc{iv}}$ for the targets of the new X-LP and for the HYPERION targets VHS~J0411-0907 and DELS~J0038-1527 which were not reported in \citetalias{Tortosa2024}. We refer the reader to \citetalias{Tortosa2024} for the corresponding measurements of the remaining targets.}
\label{table:vshift}  
\centering 
\begin{tabular}{lccc}   
\hline\hline                
Target &  $v_{\rm C\,\textsc{iv}}$ & $REW_{\rm C\,\textsc{iv}}$  & Ref.\\
& [km\,s$^{-1}$] &[$\AA$]& \\  
\hline  
\noalign{\smallskip}
VHS~J0411-0907 & $-1418\pm298 $ & $43.2\pm4.8$ & (1) \\
\hline 
DELS~J0038-1527& $-3800\pm411$ & $18.1\pm1.4$ & (1)\\
\hline
SDSS~J1411+1217  & $304$& $72$ & (2)\\
\hline 
PSO~J183-12  & $-3031 \pm 194$ & - & (3)\\
\hline
PSO~J025-11      & $-1316\pm286$& - & (3)\\
\hline
SDSS~J2054-0005 &  $-3319\pm215$ & - & (4)\\
\hline 
J0313-1806 & $-3080$ & - & (5)\\
\hline\hline
\end{tabular}
\tablebib{(1)~\citet{Yang2021};(2)~\citet{Jiang_2007};(3)~\citet{Mazzucchelli2023};(4)~\citet{Farina2022}; (5)~\citet{Wang_2021}.} 
\end{table}
\FloatBarrier
\section{Uncertainty-propagated distribution}
\label{app:gamma_distribution}
The photon index ($\Gamma$) of each source in our sample is measured with associated statistical uncertainties, which are often asymmetric due to the shape of the likelihood function in the spectral fitting. In order to characterise the uncertainty-propagated distribution of the measured $\Gamma$ values for the $z \gtrsim 6$ QSOs in our sample, and to derive robust cumulative fractions above a given threshold, it is necessary to properly propagate these uncertainties.

To this end, we adopted a Monte Carlo approach. For each source with a measured $\Gamma$, we generated a large number of realisations of $\Gamma$ (10,000) by sampling from an asymmetric Gaussian (also known as a split-normal or two-piece Gaussian) distribution. This distribution is centered on the best-fit value $\mu \equiv \Gamma_{\rm best}$ and is characterised by two different standard deviations, corresponding to the lower and upper uncertainties, $\sigma_{\rm lo}$ and $\sigma_{\rm hi}$.

The probability density function is defined as:
\begin{equation}
f(x) =
\begin{cases}
A \exp\left(-\frac{(x-\mu)^2}{2\sigma_{\rm lo}^2}\right) & x \le \mu \\
A \exp\left(-\frac{(x-\mu)^2}{2\sigma_{\rm hi}^2}\right) & x > \mu
\end{cases}
\label{eq:distribution}
\end{equation}
where $A = \sqrt{2/\pi} \, (\sigma_{\rm lo} + \sigma_{\rm hi})^{-1}$ ensures proper normalisation.

In practice, the sampling was performed by randomly selecting the side of the distribution with probabilities proportional to $\sigma_{\rm lo}$ and $\sigma_{\rm hi}$, and then drawing a value from a Gaussian distribution with the corresponding width. This procedure ensures a continuous distribution at $x=\mu$ and preserves the asymmetry of the uncertainties.

For each Monte Carlo realisation, a mock sample was constructed by drawing one value of $\Gamma$ per source. The ensemble of all simulated values was then used to reconstruct the uncertainty-propagated distribution of the measured $\Gamma$ values. This allows us to account for the impact of measurement uncertainties on the inferred properties of the sample.

From the full set of simulated $\Gamma$ values, we derived: the normalised distribution of photon indices in the population; the mean and median values of $\Gamma$, along with their associated dispersions; the 15.85th and 84.15th percentiles, corresponding to the equivalent of $\pm1\sigma$ for a Gaussian distribution without assuming symmetry; the cumulative fraction of sources with $\Gamma$ above a given threshold, defined as:

\begin{equation}
f(>\Gamma_{\rm thr}) = \frac{N(\Gamma > \Gamma_{\rm thr})}{N_{\rm tot}}
\end{equation}

where $N(\Gamma > \Gamma_{\rm thr})$ is the number of simulated values exceeding the threshold $\Gamma_{\rm thr}$, among all pooled realisations, and $N_{\rm tot}= N_{\rm src} \times 10^4 = 2.6 \times 10^5$ is the total number of simulated values ($N_{\rm src} = 26$ sources with a measured $\Gamma$, $10^4$ realisations each). This is equivalent to computing the fraction of sources with $\Gamma>\Gamma_{\rm thr}$> in each of the $10^4$ mock samples and averaging over the realisations. This quantity, reported in Fig.\ref{fig:cumulative_function}, provides a robust estimate of the fraction of QSOs with increasingly steep X-ray spectra, properly accounting for measurement uncertainties. This approach enables a statistically consistent characterisation of the $\Gamma$ distribution and avoids biases that would arise from considering only best-fit values.

The Monte Carlo procedure described above allows us to reconstruct the uncertainty-propagated distribution of $\Gamma$ for the $z \gtrsim 6$ QSOs in our sample, properly accounting for the asymmetric uncertainties on individual measurements (see Fig.\ref{fig:gamma_mc}). For comparison we reconstructed the uncertainty-propagated distribution of $\Gamma$ also for the sample of WISSH QSOs at $z\sim2-4$ from \citealt{2025A&A...702A.114D}. We restricted this comparison to the WISSH sources with a genuinely free-fitted $\Gamma$, excluding objects for which $\Gamma$ was fixed for hardness-ratio-only analysis due to limited photon statistics, to ensure a homogeneous comparison. The median, mean, 15.85th and 84.15th percentiles values of the simulated $\Gamma$ distributions of the two samples are reported in Tab.~\ref{tab:gamma_distribution}.

To quantitatively assess whether the reconstructed $\Gamma$ distributions of the $z\gtrsim6$ and WISSH QSO samples differ, we performed a two-sample Kolmogorov--Smirnov (KS) test \citep{massey1951kolmogorov}. Applying the test directly to the best-fit $\Gamma$ measurements of both samples yields a KS statistic of $D=0.60$ and a significance of $1-P=0.999$, indicating that the two samples are highly unlikely to be drawn from the same parent distribution.

To account for the measurement uncertainties, we repeated the KS test over $10\,000$ Monte Carlo realisations. At each iteration, one $\Gamma$ value was randomly drawn for each source from the corresponding split-normal distribution, and the KS statistic was recomputed. The resulting distribution of KS statistics has a median value of $D=0.51^{+0.08}_{-0.09}$, while the median significance is $1-P=0.9906$. Approximately $98\%$ of the realisations yield $1-P>0.95$, confirming that the difference between the two $\Gamma$ distributions remains statistically significant even after fully propagating the individual measurement uncertainties. We note that this difference is not driven by the different rest-frame bands probed in the two samples: restricting the spectral analysis of WISSH QSOs with comparable spectral quality to rest-frame energies above 2~keV, i.e. the same lower bound probed for our $z\gtrsim6$ sources, still yields canonical photon indices \citepalias{Zappacosta2023}.

\begin{table}
\centering
\caption{Distribution of the photon index $\Gamma$ for the two quasar samples. We report median value ($\Gamma_{\rm med}$), mean value $\langle \Gamma \rangle$, 15.85th and 84.15th percentiles ($\Gamma_{\rm 15.85}$, $\Gamma_{\rm 84.15}$) of the distribution }
\begin{tabular}{lcccc}
\hline
\hline
  & $\Gamma_{\rm med}$ & $\langle \Gamma \rangle$ & $\Gamma_{\rm 15.85}$  & $\Gamma_{\rm 84.15}$ \\
\hline
\hline
$z \gtrsim 6$   & $2.35$ & $2.40$ & 1.80 & 2.98 \\
$z\sim2-4$ & $1.90$ & $1.88$ & 1.49 & 2.21 \\
\hline
\hline
\end{tabular}
\label{tab:gamma_distribution}
\end{table}

The reconstructed uncertainty-propagated distribution of the measured $\Gamma$ values indicates that our sample is systematically characterised by steep X-ray spectra. In particular, we find that a significant fraction of sources ($\sim75\%$) exhibits photon indices above typical values observed in lower-redshift AGN (see Fig.~\ref{fig:cumulative_function}). However, it remains relatively broad due to the large uncertainties affecting the individual spectral measurements. As a consequence, the current data do not allow us to robustly assess whether an additional population of QSOs with canonical X-ray slopes is present. In such a scenario, one would expect either a bimodal distribution, with a secondary peak around $\Gamma\sim1.8-2.0$, or a skewed distribution characterised by an extended low-$\Gamma$ tail. The distribution indeed includes a non-negligible fraction ($\sim15-20\%$) of realisations with canonical slopes ($\Gamma\sim1.5-2.0$), as well as a smaller fraction ($<10\%$) with flatter spectra ($\Gamma<1.5$), potentially consistent with moderate intrinsic absorption. At the same time, the distribution also includes a significant fraction ($\geq15\%$) of very steep spectra ($\Gamma>3$).

\begin{figure}
    \centering
    \includegraphics[width=0.9\columnwidth]{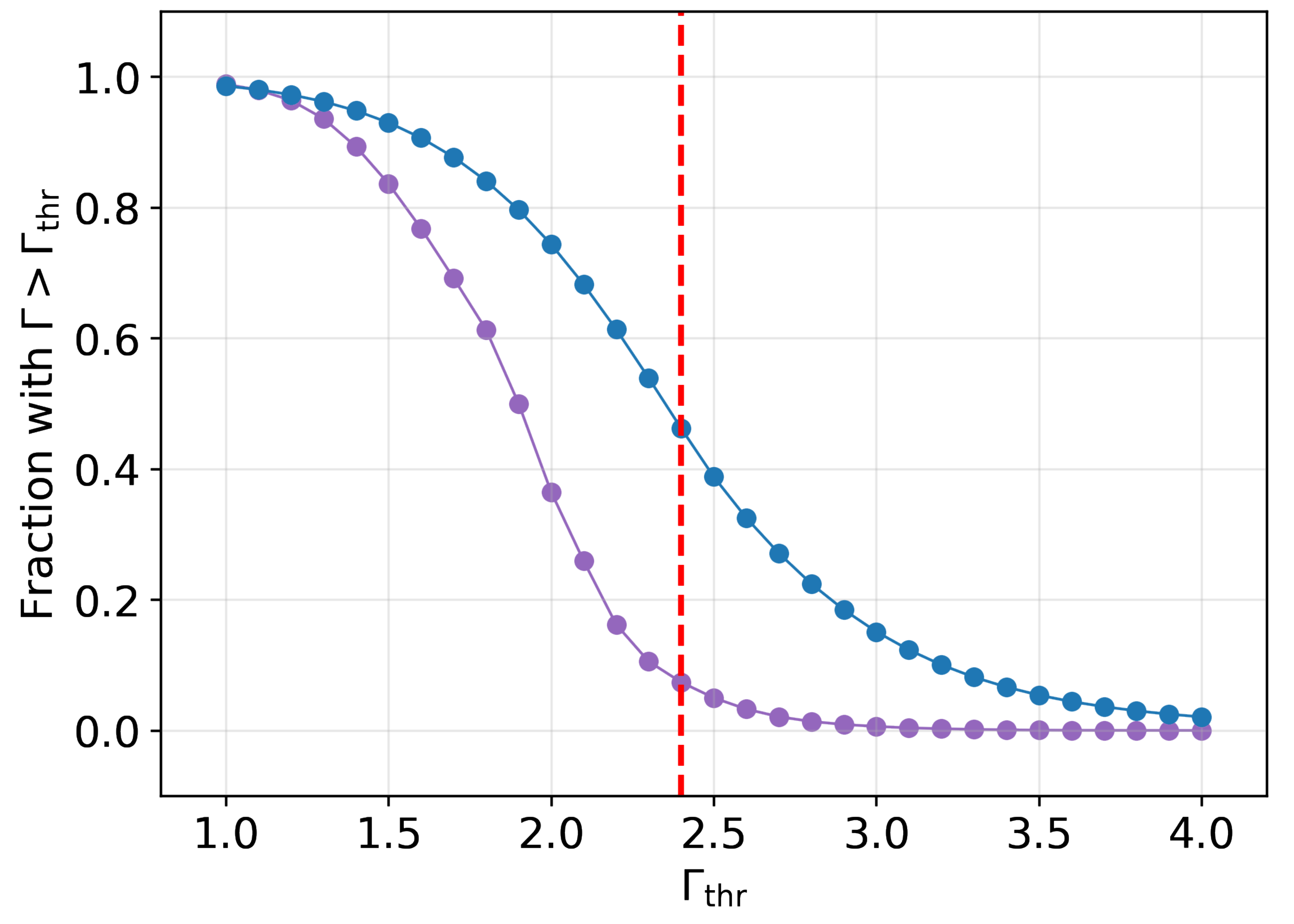}
    \caption{Cumulative fraction of sources with $\Gamma$ above a given threshold for sample of $z\gtrsim6$ QSOs analysed in this work (blue) and for the WISSH $z\sim 2-4$ QSOs sample (purple). The distribution confirms that our sample of $z \gtrsim 6$ QSOs is characterised by systematically steep X-ray spectra, with a significant fraction of sources exhibiting $\Gamma > 2.4$.}
    \label{fig:cumulative_function}
\end{figure}
These results indicate that a substantial fraction of $z \gtrsim 6$ QSOs may host X-ray coronae characterised by intrinsically steep continua, possibly supporting the scenario in which high-$z$ quasars are preferentially observed in extreme accretion states, possibly associated with high Eddington ratios and efficient cooling of the corona. Overall, the analysis demonstrates that the steepness of the X-ray spectra is a robust property of the population and not an artifact of measurement uncertainties.

\FloatBarrier
\section{Testing absorbed steep-spectrum scenarios}
\label{app:test_steep_spectra}
To investigate whether the few QSOs displaying canonical or relatively flat X-ray slopes ($\Gamma<2$) could instead be intrinsically steep-spectrum sources affected by moderate intrinsic absorption, we performed an additional spectral analysis on these objects. For each source, we first adopted the baseline spectral model used throughout this work, consisting of a Galactic absorbed power-law (\texttt{tbabs*zpowerlaw}), model A, with $\Gamma$ left free to vary and no intrinsic absorption component. We then tested a model including intrinsic absorption at the source redshift (\texttt{tbabs*ztbabs*zpowerlaw}), initially fixing the photon index to the mean value measured for the sample (i.e. $\Gamma=2.4$) while leaving the intrinsic column density free to vary, model B. Finally, we allowed both $\Gamma$ and the intrinsic column density to vary simultaneously, model C. The goodness of the fits was compared through the variation of the $\mathcal{C}$-statistic. 
In all cases, the inclusion of intrinsic absorption provides only marginal improvements in the fit quality, with typical variations of $\Delta \mathcal{C} \lesssim 2-3$ for one additional free parameter. According to the Akaike Information Criterion \citep[AIC,][]{1974ITAC...19..667A}, the absorbed models are not statistically preferred over the simpler unabsorbed power-law model ($\Delta {\rm AIC} \lesssim 1.5$, see Tab.~\ref{table:AIC}). The AIC provides a statistical method to compare models by balancing goodness of fit and model complexity, penalising the inclusion of additional free parameters. The AIC is defined as ${\rm AIC}=\mathcal{C}+2k$, where $\mathcal{C}$ is the best-fit Cash statistic and $k$ is the number of free parameters in the model. Differences of $\Delta {\rm AIC}<2$ are generally interpreted as indicating no statistically significant preference between the competing models. We note that in the low-count regime typical of high-$z$ X-ray spectra, and when using likelihood-based statistics such as the Cash statistic, AIC comparisons are generally preferred over the classical F-test, which is formally valid only for Gaussian statistics and parameters far from the boundary of the allowed parameter space.

When both $\Gamma$ and $N_{\rm H}$ are left free to vary, the best-fit $\Gamma$ remain poorly constrained and do not systematically converge toward steep values. Therefore, with the current data quality, it is not possible to establish whether the sources with canonical $\Gamma$ are intrinsically steep but moderately obscured systems, or instead represent the low-$\Gamma$ tail, or a genuine sub-population, of EoR QSOs with flatter X-ray continua.

\begin{table}\caption{Differences in the Akaike Information Criterion ($\Delta{\rm AIC}$) obtained from the comparison between the baseline unabsorbed power-law model (A) and the absorbed models with fixed (B) and free (C) photon index for the QSOs with canonical X-ray slopes ($\Gamma<2$). In all cases, $\Delta$AIC values remain below 2, indicating that the inclusion of intrinsic absorption is not statistically required by the current data.}
\label{table:AIC} 
\centering 
\begin{tabular}{cccccc}
\hline\hline
 & J0411 &J083 &J1602&J1306&J159 \\
\hline\hline
$\Delta{\rm AIC}_{\rm B-A}$ & 0.6 &0.7 &0.6&0.8&0.4 \\
$\Delta{\rm AIC}_{\rm C-A}$ & 1.2 &1.3 &1.1&1.5&1.0 \\
$\Delta{\rm AIC}_{\rm C-B}$ & 0.6 &0.6 &0.5&0.7&0.6 \\
\hline\hline
  \end{tabular}

\end{table}
\FloatBarrier
\section{Observational bias induced by spectral shape}
\label{app:bias_shape}
Here we present the full set of simulations quantifying the observational bias introduced by spectral shape (see Sect.~\ref{sec:bias_steepness}). We fix the intrinsic rest-frame 2--10~keV luminosity and compute the expected count rates for \xmm, \textit{Chandra}, and \textit{NewAthena}/WFI as a function of (i) photon index $\Gamma$ and (ii) high-energy cutoff $E_{\mathrm c}$. For each configuration we show the bias as the ratio between the predicted count rate and that obtained assuming a canonical spectral model ($\Gamma=1.9$ or $E_{\mathrm c}=150$ keV). The simulations are shown for different redshifts from $z=5$ to $z=10$ in the observed 0.3--2~keV, 0.3--7~keV and 2--10~keV bands. Fig.~\ref{fig:bias_gamma} and Fig.~\ref{fig:bias_Ec} show how steep spectra can significantly suppress the observed count rate, particularly in the hard observed band, even when the intrinsic 2--10~keV luminosity is kept fixed.
\begin{figure*}
    \centering
    \includegraphics[width=\textwidth]{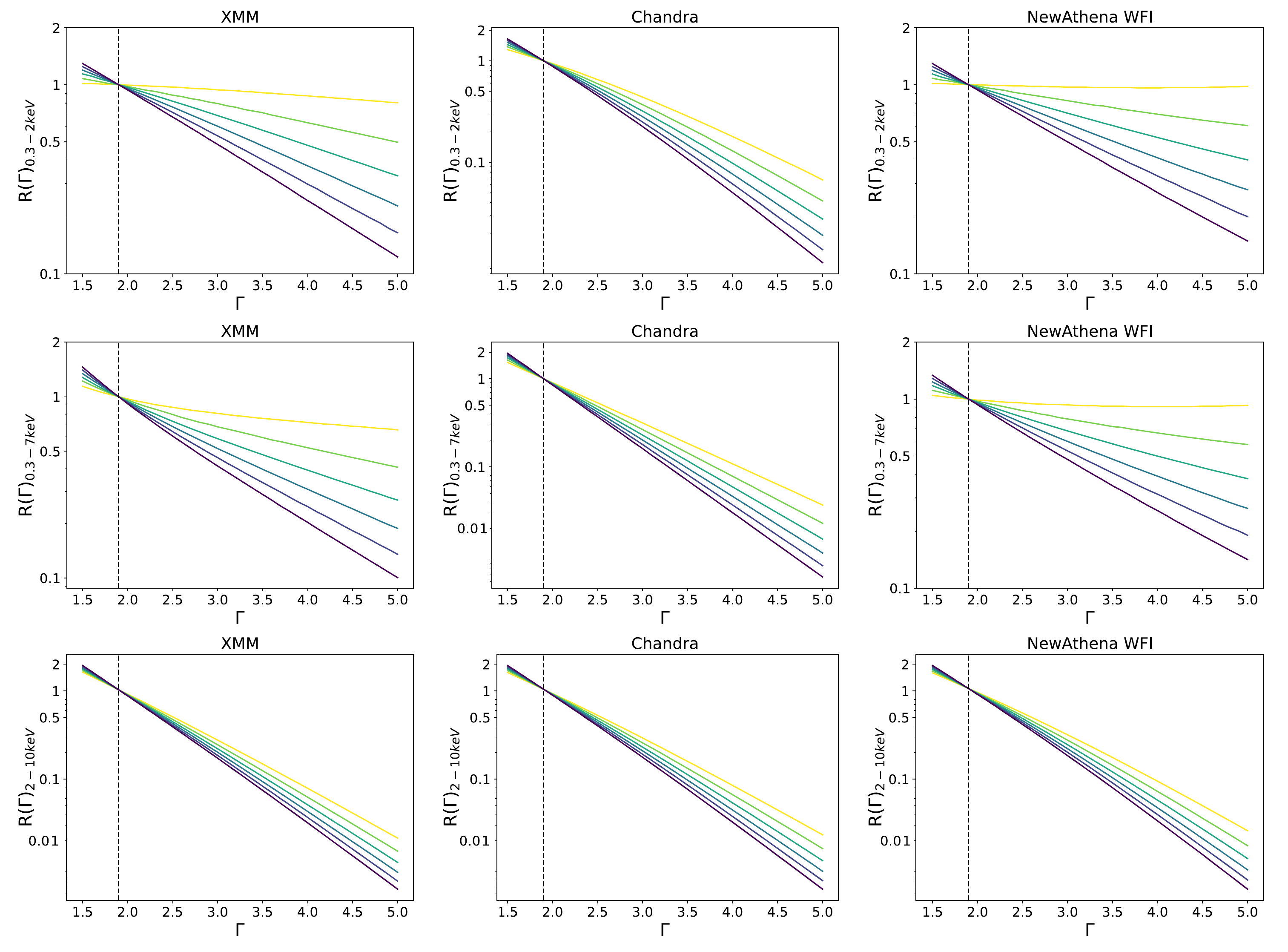}
    \caption{The "spectral steepness bias" ($R(\Gamma)_{EB}$) as a function of photon index ($\Gamma$), for \textit{XMM-Newton}, \textit{Chandra},  and \textit{NewAthena}/WFI. Count rates ratios in the 0.3--2~keV, 0.3--7~keV and 2--10~keV bands are shown in top panels, middle panels and bottom panels, respectively. Different curves correspond to increasing redshift from $z=5$ (yellow) to $z=10$ (dark purple). The intrinsic rest-frame 2--10~keV luminosity is held fixed.}
    \label{fig:bias_gamma}
\end{figure*}

\begin{figure*}
    \centering
    \includegraphics[width=\textwidth]{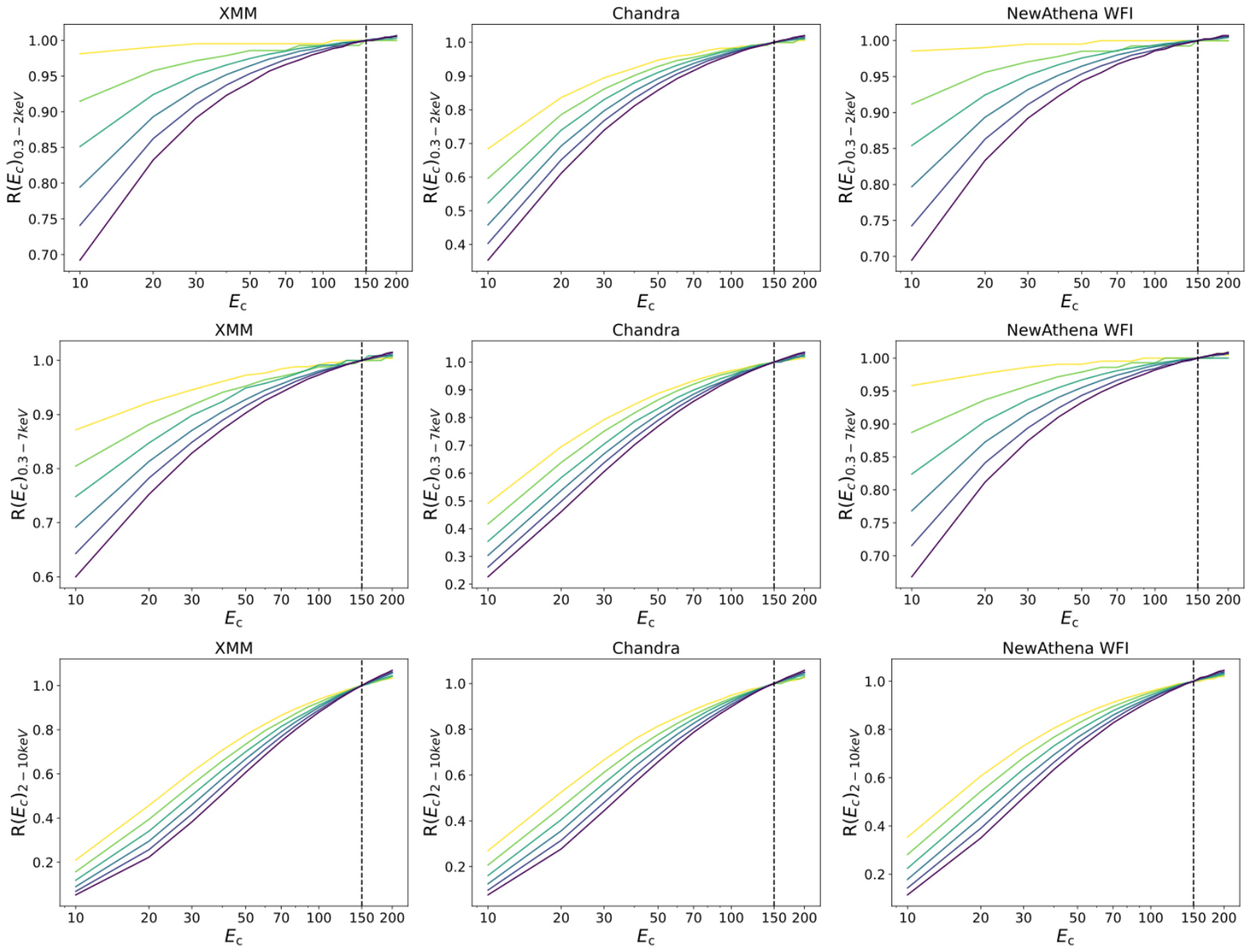}
    \caption{The "spectral cutoff energy bias" ($R(E_{\rm c})_{\rm EB}$) as a function of high-energy cutoff ($E_{\rm c}$), for \textit{XMM-Newton}, \textit{Chandra}, and \textit{NewAthena}/WFI. Count rates ratios in the 0.3--2~keV, 0.3--7~keV and 2--10~keV bands are shown in top panels, middle panels and bottom panels, respectively. Different curves correspond to increasing redshift from $z=5$ (yellow) to $z=10$ (dark purple). The intrinsic rest-frame 2--10~keV luminosity is held fixed.}
    \label{fig:bias_Ec}
\end{figure*}
\FloatBarrier
\end{appendix}
\end{document}